\documentclass[aps,nofootinbib,floatfix,showpacs,preprintnumbers,amsmath,amssymb]{revtex4}
\makeatletter
\def\l@subsubsection#1#2{}
\makeatother
\usepackage[colorlinks]{hyperref}
\usepackage{natbib}
\usepackage{dcolumn}
\usepackage{enumerate}
\usepackage{multirow}
\usepackage{inputenc}
\usepackage{fontenc}
\usepackage{float}
\usepackage{graphicx, rotating, lscape, threeparttable}
\usepackage[colorlinks]{hyperref}
\hypersetup{%
  colorlinks = true,
  linkcolor  = black
}
\usepackage{changepage}
\newcommand{\be}{\begin{equation}}
\newcommand{\ee}{\end{equation}}
\newcommand{\bea}{\begin{eqnarray}}
\newcommand{\eea}{\end{eqnarray}}

\def\[{\begin{equation}}
\def\]{\end{equation}}
\begin{document}
\title{\texttt{ISiTGR}: Testing deviations from GR at cosmological scales including dynamical dark energy, massive neutrinos, functional or binned parametrizations, and spatial curvature}
\date{\today}
\author{Cristhian Garcia-Quintero}
\email{gqcristhian@utdallas.edu}
\author{Mustapha Ishak}
\email{mishak@utdallas.edu}
\author{Logan Fox}
\author{Jason Dossett}
\affiliation{Department of Physics, The University of Texas at Dallas, Richardson, Texas 75080, USA}
\begin{abstract}
We introduce a new version of the {\it {I}ntegrated {S}oftware {i}n {T}esting {G}eneral {R}elativity} (\texttt{ISiTGR}) which is a patch to the software \texttt{CAMB} and \texttt{CosmoMC}. \texttt{ISiTGR} is intended to test deviations from GR at cosmological scales using cosmological data sets. While doing so, it allows for various extensions to the standard flat $\Lambda$CDM model. In this new release, we have support for the following:  
1) dynamical dark energy parametrizations with a constant or time-dependent equation of state; 2) a consistent implementation of anisotropic shear to model massive neutrinos throughout the full formalism; 
3) multiple commonly-used parametrizations of modified growth (MG) parameters; 
4) functional, binned and hybrid time- and scale-dependencies for all MG parameters; 
5) spatially flat or curved backgrounds. 
\texttt{ISiTGR} is designed to allow cosmological analyses to take full advantage of ongoing and future surveys to test simultaneously or separately various extensions to the standard model. We describe here the formalism and its implementation in the CMB code, the Integrated Sachs-Wolfe (ISW) effect, and the 3x2 point statistics.  Next, we apply \texttt{ISiTGR} to current data sets from Planck-2018, Planck-2015, Dark Energy Survey YR1 release, Baryonic Acoustic Oscillations (BAO), Redshift Space Distortions (BAO/RSD) from the BOSS Data Release 12, the 6DF Galaxy Survey and the SDSS Data Release 7 Main Galaxy Sample, and Supernova from the Pantheon compilation, joint SNLS/SDSS data analysis and the Hubble Space Telescope.
We derive constraints on MG parameters for various combinations of the five features above and find that GR is consistent with current data sets in all cases. The code is made publicly available at \url{https://github.com/mishakb/ISiTGR}.
\end{abstract}
\maketitle
\newpage
\setcounter{tocdepth}{2}
\tableofcontents
\newpage

%%%%%%%%%%%%%%%%%%%%%%%%%%%%%%%%%%%%%%%%%%%%%%%%%

%%%%%%%%%%%%%%%%%%%%%%%%%%%%%%%%%%%%%%%%%%%%%%%%%
\section{Introduction}
%%%%%%%%%%%%%%%%%%%%%%%%%%%%%%%%%%%%%%%%%%%%%%%%%%%%
%
Continuous progress is being made toward precision cosmology with a number of ongoing and planned surveys and missions \cite{KIDSws,DESws,HSCws,LSSTws,DESIws,Euclidws,SKAws,WFIRSTws,SIMONws,CMB-S4}$^1$ \footnotetext[1]{
e.g. Kilo-Degree Survey (KiDS) \cite{KIDSws}, 
Dark Energy Survey (DES) \cite{DESws}, 
Hyper Suprime-Cam (HSC) \cite{HSCws}, 
Large Synoptic Survey Telescope (LSST) \cite{LSSTws}, 
Dark Energy Spectroscopic Instrument (DESI) \cite{DESIws}, 
Euclid \cite{Euclidws},
Square Kilometre Array (SKA) \cite{SKAws}, 
Wide Field Infrared Spectroscopic Telescope (WFIRST) \cite{WFIRSTws}, 
Simons Observatory \cite{SIMONws}, CMB-S4 \cite{CMB-S4} and many others.}  
The resulting complementary and precise observations have opened the door to testing gravity physics (General Relativity (GR)) at cosmological scales, see, e.g., the reviews \cite{2012-Clifton-MG,KOYAMA2016TestGR,2015-rev-Joyce-et-al,2016-Joyce-Lombriser-Schmidt-DEvsMG,IshakMG2019}. 

One of the chief motivations to test GR at cosmological scales is the pressing question of cosmic acceleration and the dark energy associated with it, see e.g. \cite{Weinberg1988CC,Carroll2001CC,Sahni2000,Peebles2003,2007-Ishak-Remarks-DE,Weinberg2013,Huterer2018}. However, testing GR at large scales is a well-motivated and justified objective in its own right.

There are, at least, two primary routes to testing gravity at large scales. The first is to model departure from GR in a phenomenological way by adding parameters that would signal such a deviation at the level of the growth of large-scale structure in the Universe. Such parameters would take some expected values in GR, often one or zero, but will deviate from them otherwise. Interestingly, different models of gravity that have the same expansion history can still exhibit distinct growth rates of large-scale structures, which can be used as a discriminant between gravity theories, see e.g. \cite{Linder2005,
2006-Koyama-growth-in-MG,ZhangEtAl2007,CaldwellEtAl2007,LinderCahn2007,Polarski2008,
GongCurved2009,ZhaoEtAl2009,Acquaviva2010,BeanTangmatitham2010,Lombriser2011,DossettFOM2011}.
 The second approach is to develop analysis pipelines and simulations specific to some proposed modified gravity theories such as the well-known $f(R)$, the DGP (Dvali-Gabadadze-Porrati) \cite{DGP}, or other modified gravity models. Understandably, this second approach has been progressing at a slower  pace because it requires more involved development and  resources. The two methods complement each other in the effort to test gravity at cosmic scales. For further discussion on both approaches, we refer the reader to the following partial list of reviews  \cite{2012-Clifton-MG,KOYAMA2016TestGR,2015-rev-Joyce-et-al,2016-Joyce-Lombriser-Schmidt-DEvsMG,IshakMG2019} and references therein. In view of recent interest and  developments, it is also worth mentioning the method of exploring inconsistencies between data sets as a way to test departures from the standard model, see e.g.  \cite{2006-Ishak-splitting,2015Ruiz-etal-param-splitting,2016-Bernal-Verde-Riess-H0,Lin2017A,GQC2019}.

In this paper, we introduce and describe a new version of (\texttt{ISiTGR}) ({\it \textbf{I}ntegrated \textbf{S}oftware \textbf{i}n \textbf{T}esting \textbf{G}eneral \textbf{R}elativity})  \cite{ISITGR,Dossett2012}  which is a patch for the widely-used software packages \texttt{CAMB} (Code for Anisotropies in the Microwave Background  \citep{CAMB})  and \texttt{CosmoMC} (Cosmological Monte Carlo \citep{COSMOMC}). \texttt{ISiTGR} follows the first approach described above to constrain departure from GR based on various pairs of modified growth (MG) parameters. These parametrize the strength of the coupling between the gravitational potentials and the spacetime sources, as well as the relationships between the two potentials, as we describe further below in the paper.    

In this new release of \texttt{ISiTGR}, we included contributions from anisotropic shear stress throughout all of the formalism to consistently account for contributions from massive neutrinos and radiation. We also included dynamical dark energy for the background with constant or time-varying equations of state. 
Furthermore, we expanded the support for several pairs of existing MG parametrizations as needed by various types of cosmological probes and surveys. These features have been made to work consistently in a spatially flat or curved background. Additionally, we implement functional as well as binned methods for the time and scale dependencies of MG parameters. In this way, the new version of \texttt{ISiTGR} has been designed to suit the needs of analyses that intend to test various aspects of extended models using incoming and future data sets,  and makes it possible to constrain such extensions separately or \emph{simultaneously}. \texttt{ISiTGR} has been cited or used in over 50 papers and has been applied to CFHTLens, KidS-450, 2dF and Planck data, see for example \cite{JoudakiEtAl2017,Joudaki2018,Dossett2015}.
 
There are over a dozen other codes that test deviation from GR or specific MG models at cosmological scales. These include for example  
\texttt{MGCAMB} \cite{ZhaoEtAl2009,MGCAMB3} that is similar to \texttt{ISiTGR} and is built on the top of \texttt{CAMB}; hi\_class  \cite{HICLASS} that is built on CLASS \cite{CLASS-I,CLASS-II} and based on the Horndeski models (and beyond) \cite{Horndeski1974}; \texttt{EFTCAMB} \cite{EFTCAMB1,EFTCAMB2} which follows an approach inspired by Effective Field Theory perturbations applied to dark energy; and \texttt{EoS\_class} that is based on the equation of state approach applied to Horndeski models \cite{EOSCLASS2019}. We refer the reader for overviews of codes in  Refs. \cite{BelliniEtAl2017AC,IshakMG2019}.   

The paper is organized as follows. In section II, we describe the growth equations in a flat or curved background with anisotropic shear stress. Effective dynamical dark energy evolution with constant or time-varying equations of state is summarized in section III. In section IV, we describe modified growth (MG) equations and various MG parameters as used in various surveys and probes. We describe there as well the time and scale dependencies for MG parameters as implemented in functional and binned forms. Section V is a brief overall description of how the modifications have been implemented in the code, while section VI describes power spectra and correlation functions for the observables. In section VII, we apply \texttt{ISiTGR} to current available data sets to constrain MG parameters in different forms and  backgrounds. We summarize in section VIII. 

% MI??? ????Examples of this approach can be seen in \cite{r17,r53,r35, r33, r54, r34, r36, r37, r38, r55, r39, r40, r56, r57, r41} to name a few.  

%%%%%%%%%%%%%%%%%%%%%%%%%%%%%%%%%%%%%%%%%%%%%%%%%%%%%%%%%%%%%%%%%%%%%%%
\section{Growth equations including anisotropic shear and spatial curvature}
%%%%%%%%%%%%%%%%%%%%%%%%%%%%%%%%%%%%%%%%%%%%%%%%%%%%%%%%%%%%%%%%%%%%%

%%%%%%%%%%%%%%%%%%%%%%%%%%%%%%%%%%%%%%%%%%%%%%%%%%%%%%%
\subsection{Growth Equations}
%%%%%%%%%%%%%%%%%%%%%%%%%%%%%%%%%%%%%%%%%%%%%%%%%%%%%%%
 
As discussed above, a modification to GR at cosmological scales can affect the growth rate of large scale structure. This can be phenomenologically modeled by changing the linearly perturbed Einstein equations. Specifically, one can change the resulting Poisson-like equations involving the gravitational scalar potentials. Let us first briefly review the key equations in the GR case where we allow for spatial hyper-surfaces to be flat or curved and we keep the anisotropic shear stress coming from, for example, the Neutrino sector. 

We start with the perturbed Friedmann–Lema\^{î}tre–Robertson–Walker (FLRW) metric written in the general conformal Newtonian gauge given by
\be
ds^2=a(\tau)^2[-(1+2\Psi)d\tau^2+(1-2\Phi)\gamma_{ij}dx^i dx^j],
\label{eq:FLRWNewt}
\ee
where $\Phi$ and $\Psi$ are scalar gravitational potentials describing the scalar mode of the metric perturbations, $x_i$'s are the comoving coordinates, $\tau$ is conformal time, and $a(\tau)$ is the scale factor. $\gamma_{ij}$ is the 3-dimensional metric, which can be written in the Cartesian coordinates $(x,\,y,\,z)$ as \cite{Mueller2009}
\be
\gamma_{ij} = \delta_{ij}\left[1+\frac{K}{4}\left(x^2+y^2+z^2\right)\right]^{-2},
\label{eq:3met}
\ee
where $K=-\Omega_k\mathcal{H}_0^2$ is the spatial curvature and we use units where $c=1$. $\mathcal{H}_0$ is the Hubble parameter (constant) today. 

In a non-flat FLRW Universe the Fourier modes can be generalized as eigen-functions, $G$, of the Laplacian operator such that $\nabla^2G(\vec{k},\vec{x})= -k^2G(\vec{k},\vec{x})$. One can then expand perturbations in terms of the eigen-function $G$ and its spatial covariant derivatives, see e.g. \cite{Abbott1986,Zaldarriaga1998}.

The first order perturbed Einstein equation gives two equations. The first equation { derives from the combination of the time-space and time-time perturbed equations and gives} a Poisson equation relating the potential, $\Phi$, and the gauge-invariant, rest-frame over-density, $\Delta_i$. The second equation derives from the traceless, space-space component of the equations and gives a relation between the two metric potentials involving the shear stress, $\sigma_i$ (where $_i$ denotes a particular matter species). The two equations read: 
\bea
\left(k^2-3K\right)\Phi  &=&-4\pi G a^2\sum_i \rho_i \Delta_i,
\label{eq:Poisson}\\
k^2(\Psi-\Phi) &=& -12 \pi G a^2\sum_i \rho_i(1+w_i)\sigma_i,
\label{eq:2ndEin}
\eea
where $\rho_i$ is the density for matter species $i$.  

The gauge-invariant, rest-frame overdensity, $\Delta_i$,
is a key quantity since its evolution describes the growth of inhomogeneities (structures) in the Universe. It is defined as
\be
\Delta_i = \delta_i +3\mathcal{H}\frac{q_i}{k},
\label{eq:Delta}
\ee 
where $\mathcal{H} =\dot{a}/a$ is the Hubble factor in conformal time; $\delta_i=\delta \rho_i/\bar{\rho}$ is the fractional overdensity;  and $q_i$ is the heat flux related to the divergence of the peculiar velocity, $\theta_i$, by $\theta_i=\frac{k\ q_i}{1+w_i}$.

From conservation of the energy momentum tensor of the perturbed matter fluids, the fractional overdensity and heat flux for uncoupled fluid species or the mass-averaged quantities for all the fluids evolve as \cite{Ma-Bertschinger1995}:
\bea
\dot{\delta} & = & -k q +3(1+w)\dot{\Phi}+3\mathcal{H}(w-\frac{\delta P}{\delta\rho})\delta,
\label{eq:deltaevo}\\
\frac{\dot{q}}{k}&= &-\mathcal{H}(1-3w)\frac{q}{k}+\frac{\delta P}{\delta\rho}\delta+(1+w)\left(\Psi-\sigma\right),
\label{eq:qevo}
\eea
where $w=p/\rho$ is the equation of state of the fluid. Next, combining the two equations above, one can express the evolution of $\Delta$ (or $\Delta_i$) by
\be
\dot{\Delta} = 3(1+w)\left(\dot{\Phi}+\mathcal{H}\Psi\right)+3\mathcal{H}w\Delta -\left[k^2+3\left(\mathcal{H}^2-\dot{\mathcal{H}}\right)\right]\frac{q}{k}-3\mathcal{H}(1+w)\sigma.
\label{eq:Deltadot}
\ee
Combining the growth equations (\ref{eq:Poisson}) and (\ref{eq:2ndEin}), along with the evolution equations (\ref{eq:Deltadot}) and $a(\tau)$, the growth history of large scale structures in the Universe can be fully described.

\subsection{Massive Neutrino contributions}

When considering consistently the contributions from massive neutrinos to MG equations, one needs to include the anisotropic shear stress throughout the evolution equations. In \texttt{CAMB} it is defined as
\begin{equation}
\Pi_i = \frac{3}{2}(1+w_i)\sigma_i.
\label{Stress}
\end{equation}
In previous implementations of MG codes \cite{ZhaoEtAl2009,ISITGR}, the anisotropic shear stress contributions related with $\dot{\Pi}$ for massive neutrinos were usually neglected. 
In view of recent interest and developments in the neutrino sector, 
we have consistently included the anisotropic shear stress contributions at all levels and calculations involving MG modifications. These are described in section \ref{sec:Implementation}. 

\section{Effective Dynamical Dark Energy parametrization and evolution}
\label{sec:EffectiveDarkEnergy}

We model further below the time evolution of MG parameters via the effective dark energy density parameter's dependence on the scale factor. This is given by:
\begin{equation}
\Omega_{DE}(a) \equiv \rho_{DE} (a)/\rho_c (a).
\end{equation}
Once the effective equation of state for the dark energy is specified, we can compute $\rho_{DE} (a)$, and thus obtain an expression for $\Omega_{DE}(a)$:

\begin{equation}
\Omega_{DE}(a) = \Omega_\Lambda \left(\frac{H_0}{H}\right)^2 \exp \left[-3\int_{a_0}^{a} [1+w(a')] \frac{da'}{a'}\right].
\label{DEmodel}
\end{equation}
We solve for this dark energy parameter for four different models: standard $\Lambda$CDM model; $w$CDM model, the Chevallier-Polarski-Linder (CPL) parametrization (see \cite{Chevallier2001IAU,Linder2003ET}); and the pivot equation of state for the CPL parametrization described in \cite{Albrecht2006}. The corresponding solutions to Eq. (\ref{DEmodel}) for each of these models along with the corresponding dark energy densities are given in Table (\ref{TableDEnergy}).
\begin{table}[b]
%\begin{adjustwidth}{-1cm}{}
\begin{tabular}{|c|c|c|}
\hline\hline
Dark Energy model &  Dark Energy density & Dark Energy evolution \\ \hline
& & \\
$w=-1$ & $\rho_v (t) = \rho_v^{(0)}$ & $\Omega_{DE}(a)=\Omega_v \left( \frac{H_0}{H}\right)^2$ \\
& & \\ 
$w=w_0$ & $\rho_v (t) = \rho_v^{(0)} a^{-3(1+w_0)}$ & $\Omega_{DE}(a)=\Omega_v \left( \frac{H_0}{H}\right)^2 a^{-3(1+w_0)}$ \\
& & \\ 
$w(a)=w_0+(1-a)w_a$ & $\rho_v (t) = \rho_v^{(0)} a^{-3(1+w_0+w_a)} e^{3 w_a (a-1)} $ & $\Omega_{DE}(a)=\Omega_v \left( \frac{H_0}{H}\right)^2 a^{-3(1+w_0+w_a)} e^{3w_a(a-1)}$ \\
& & \\ 
$w(a)=w_p+(a_p-a)w_a$ & $\rho_v (t) = \rho_v^{(0)} a^{-3(1+w_p+a_p w_a)} e^{3 w_a (a-1)} $ & $\Omega_{DE}(a)=\Omega_v \left( \frac{H_0}{H}\right)^2 a^{-3(1+w_p+a_p w_a)} e^{3w_a(a-1)} $ \\
& & \\ 
\hline\hline
\end{tabular}
\caption{Dark energy parametrizations given in \texttt{ISiTGR}: $\Lambda$CDM, $w$CDM, the ($w_0$,$w_a$) parametrization and a pivot dark energy equation of state ($w_p$,$w_a$). Once the user chooses one of these models, \texttt{ISiTGR} selects the corresponding dark energy density and the corresponding dark energy evolution.}
\label{TableDEnergy}
%\end{adjustwidth}
\end{table}

We note that a pivot dark energy equation of state based on the CPL parametrization is implemented in \texttt{ISiTGR}. This is done through a linear transformation giving $w=w_p+(a_p-a)w_a$, where $a_p$ is the scale factor related to a pivot redshift $z_p=1/a_p -1$. The pivot scale factor is determined by \cite{Albrecht2009}
\begin{equation}
a_p=1+\boldsymbol{C}(w_0,w_a)/\boldsymbol{C}(w_a,w_a),
\end{equation}
where $\boldsymbol{C}$ represents the covariance matrix. The choice of $a_p$ is such that $w_p$ and $w_a$ are decorrelated so it minimizes the uncertainty in $w(a)$, but  user should be aware of some limitations or misinterpretations expressed in, e.g. see \cite{Linder2006}.

%%%%%%%%%%%%%%%%%%%%%%%%%%%%%%%%%%%%%
%%%%%%%%%%%%%%%%%%%%%%%%%%%%%%%%%%%%%
%%%%%%%%%%%%%%%%%%%%%%%%%%%%%%%%%%%%%
\section{Modified Growth Equations and MG parameters including time and scale dependencies}
\label{sec:MGP}

As usual, we implement deviations from GR using modifications of the linearly perturbed Einstein equations and the resulting modified versions of the metric potential equations (\ref{eq:Poisson}) and (\ref{eq:2ndEin}). Again, to implement \texttt{ISiTGR} for both spatially flat and curved backgrounds, we do not limit the formalism to the $K=0$ -case (However, the explicit equations for the flat case are given in Appendix \ref{sec:flatcase}). {We also keep the shear terms non-zero throughout the equations.} Further below we provide the  functional dependencies on time (scale factor) and scale, and allow for a time-dependent equation of state for dark energy.   

\subsection{The $(\mu(a,k),\,\,\gamma(a,k))$ parametrization}
The first parameter $(\mu(a,k))$ enters the two MG equations below, (\ref{muEquation}) and (\ref{gammaEquation}), and quantifies the strength of the gravitational coupling between the potentials and the sources. 
The second parameter, $\gamma(a,k)$, (sometimes also noted as $\eta(a,k)$), is called the slip parameter \cite{CaldwellEtAl2007} and quantifies the difference between the two gravitational potentials. At late times when anisotropic shear can be assumed to be negligible then $\gamma(a,k) \equiv \Phi/\Psi$. The modified equations read:
\begin{equation}
(k^2-3K)\Psi = -4\pi G a^2\mu(a,k) \sum_i\left[\rho_i\Delta_i+3\left(\frac{k^2-3K}{k^2}\right)\rho_i(1+w_i)\sigma_i\right]
\label{muEquation}
\end{equation}
and
\begin{equation}
k^2(\Phi-\gamma(a,k)\Psi) = 12\pi G a^2 \mu(a,k) \sum_i\rho_i(1+w_i)\sigma_i.
\label{gammaEquation}
\end{equation}
GR is recovered when $\mu(a,k)$ and $\gamma(a,k)$ are equal to unity. This parametrization has been used in for example \cite{Planck2015MG,Planck2018} with evolutions that we describe further below. 

%%%%%%%%%%%%%%%%%%%%%%%%%%%%%%%%%%%%%%%%%%
\subsection{The $(\mu(a,k),\,\,\Sigma(a,k))$ parametrization}
{Some observables such as gravitational lensing or the Integrated Sachs-Wolfe (ISW) effect involve the Weyl potential, 
\begin{equation}
\Phi_{W}=(\Phi+\Psi)/2,
\label{eq:WeylPotential}
\end{equation}
which governs the motion of light-like particles. 
 Taking advantage of this, (\ref{eq:Poisson}) and (\ref{eq:2ndEin}) can be combined, and another parameter, $\Sigma(a,k)$, can be defined to directly probe modifications to the Weyl potential} (see e.g. \cite{SimpsonEtAl2013}). Consequently, this parameter can be defined through
\begin{equation}
k^2(\Phi+\Psi) = -4\pi G a^2\Sigma(a,k) \sum_i\left[\frac{2\rho_i\Delta_i}{1-3K/k^2} + 3\rho_i(1+w_i)\sigma_i\right].
\label{SigmaEquation}
\end{equation}
Deviation from GR is now measured using Eqs. (\ref{muEquation}) and (\ref{SigmaEquation}). Moreover, $\Sigma(a,k)$ also takes the value of unity in the GR case. In the case of negligible or zero shear, it follows that  
\be
\Sigma (k,a)= \frac{{\mathit{\mu}}(k,a)[1+{\mathit{\gamma}}(k,a)]}{2}.
\label{eq:Sigma2}
\ee
This parametrization  or a similar one have been used in, for example, \cite{SimpsonEtAl2013,Dossett2015}.

%%%%%%%%%%%%%%%%%%%%%%%%%%%%%%%%%%%%%%%
\subsection{The $(Q(a,k),\,\,R(a,k),\,\,D(a,k))$ parametrization}
Another parametrization similar to the above was introduced in \cite{BeanTangmatitham2010} with a parameter $Q(a,k)$ to characterize the gravitational strength and $R(a,k)$ as a gravitational slip parameter. The MG equations read: 
\begin{equation}
(k^2-3K)\Phi = -4\pi G a^2 Q(a,k) \sum_i\rho_i\Delta_i
\label{QEquation}
\end{equation}
and
\begin{equation}
k^2(\Psi-R(a,k)\Phi) = -12\pi G a^2 Q(a,k) \sum_i\rho_i(1+w_i)\sigma_i.
\label{REquation}
\end{equation}
Similarly, at late times when anisotropic stress is negligible, $\Psi = R\Phi$.

Again, it is also possible to combine the two equations above and define the parameter $D(a,k) = Q(a,k)(1+R(a,k))/2$ to be used instead of $R(a,k)$. {This not only avoids a strong degeneracy between $Q(a,k)$ and $R(a,k)$, but also gives a parameter which can be directly probed by lensing and ISW observations.}  The combined equation reads
\begin{equation}
k^2(\Phi+\Psi) = \left(\frac{-8\pi G a^2}{1-3K/k^2}\right) D(a,k) \sum_i \rho_i\Delta_i - 12\pi G a^2 Q(a,k)\sum_i\rho_i(1+w_i)\sigma_i.
\label{DEquation}
\end{equation}
Then, for this parametrization, one can use (\ref{QEquation}) and (\ref{DEquation}) instead of (\ref{REquation}). 
We implemented both pairs in \texttt{ISiTGR}. 

{Here it is important to note that $\Sigma (a,k)= D(a,k)$ only in the zero anisotropic shear stress case, and that while $D(a,k)\equiv Q(a,k)(1+R(a,k))/2$ is a definition, the relationship $\Sigma(a,k) = \mu(a,k)(1+\gamma(a,k))/2$ also holds only in the zero anisotropic shear case.}

%%%%%%%%%%%%%%%%%%%%%%%%%%%%%%%%%%%%%%%%%%
\subsection{MG parameter functional dependencies in time and scale}
\label{subsec:MGparamz}
We implement in \texttt{ISiTGR} the dependencies of each set of MG parameters on both time and scale. For the time evolution of MG parameters, we mainly use the dark energy density time evolution, described by the parameter $\Omega_{DE}(a)$, as shown further below for various pairs of parameters. In this way, the contribution to clustering and anisotropic stress by MG effects is proportional to their effective dark energy density. One needs to ensure though that the time evolution of the effective dark energy density is governed by its corresponding effective equation of state, as we describe in section \ref{sec:EffectiveDarkEnergy}. Note that some limitations and failures of this proportionality of MG parameters with the parameter $\Omega_{DE}(a)$ have been reported and discussed in \cite{Linder2017OmegaDE1} and \cite{Linder2017OmegaDE2}. It was reported there that such parametrization can miss deviations from GR and the proportionality does not hold for all cosmic evolution including the epoch of interest of cosmic acceleration. Then, the user should use these with some caution and complement them with binning methods.  

It was shown in \cite{DossettDEspeed} and other works that scale dependence can provide further insights to discriminate between gravity theories. For this dependence, we implemented the function used in, for example, \cite{Planck2015MG} and \cite{BZ2008}, with the additional factor 
\begin{equation}
S_{i}(a,k) = \frac{1+c_i \left( \lambda H(a)/k \right)^2}{1+\left( \lambda H(a)/k \right)^2},
\label{Sfunction}
\end{equation}
where the index $i=1,2$ stands for the first and the second MG parameters and $H(a)$ is the Hubble parameter. The form of $S_{i}(a,k)$ is such that at small scales (large $k$) we have $S_{i}\rightarrow 1$, while at large scales (small $k$) we get $S_{i} \rightarrow c_i$. Therefore, at small scales (\ref{Sfunction}) has no effect on the MG parameters, while for large scales it makes MG parameters proportional to $c_1$ and $c_2$. Therefore $S_{i}$ provides information about how MG parameters evolve at large scales. Moreover, each MG parameter may have a different scale dependent evolution, since $c_1$ is independent of $c_2$ (recovering GR when $c_1=c_2=1$, regardless of the value of $\lambda$). However, even if $c_i \sim 0$, the $\lambda$ parameter will still provide some scale dependence. In this case, we have that $S_{i}\rightarrow 1$ for large $k$,  but for small $k$ we observe that $S_{i}\rightarrow 0$. Therefore, if $c_i$ is negligible, then at large scales the MG parameters will go to their GR limit value, but at small scales the MG parameters still evolve in time. Furthermore, we can recover GR if $\lambda=0$ regardless of the values that $c_1$ and $c_2$ can take.

Next, we apply these functional dependencies to each set of MG parameters. 

%%%%%%%%%%%%%%%%%%%%%%%
\subsubsection{Time and scale functional dependencies for $(\mu(a,k)$, $\gamma(a,k))$}
\noindent For the ($\mu$, $\gamma$) case, we implemented the parametrization as given in the recent Planck analyses \cite{Planck2015MG,Planck2018}. The MG parameters are implemented explicitly as
\begin{equation}
\mu(a,k)=1+E_{11}\Omega_{DE}(a) \left[ \frac{1+c_1 \left( \lambda H(a)/k \right)^2}{1+\left( \lambda H(a)/k \right)^2} \right]
\label{muEvolutionPlanck}
\end{equation}
and
\begin{equation}
\gamma(a,k)=1+E_{22}\Omega_{DE}(a) \left[ \frac{1+c_2 \left( \lambda H(a)/k \right)^2}{1+\left( \lambda H(a)/k \right)^2} \right].
\label{gammaEvolutionPlanck}
\end{equation}
As we mentioned before, the scale dependent implementation is such that, for small $k$ we have that $\mu \rightarrow 1+c_1 E_{11}\Omega_{DE}$ and $\gamma \rightarrow 1+c_2 E_{22}\Omega_{DE}$. Furthermore, for large $k$ we note that $\mu \rightarrow 1+E_{11}\Omega_{DE}$ and $\gamma \rightarrow 1+E_{22}\Omega_{DE}$. In all cases, $\Omega_{DE}$ becomes negligible at early times (high redshift) so MG parameters go to the GR value of 1. So MG effects here are modeled to be negligible at early times. 

%%%%%%%%%%%%%%%%%%%%%%%
\subsubsection{Time and scale functional dependencies for $(\mu(a,k)$, $\Sigma(a,k))$}
We implemented here the following time and scale dependencies:
\begin{equation}
\mu(a,k)=1+\mu_{0}\frac{\Omega_{DE}(a)}{\Omega_\Lambda} \left[ \frac{1+c_1 \left( \lambda H(a)/k \right)^2}{1+\left( \lambda H(a)/k \right)^2} \right]
\label{muEvolution}
\end{equation}
and
\begin{equation}
\Sigma(a,k)=1+\Sigma_{0}\frac{\Omega_{DE}(a)}{\Omega_\Lambda} \left[ \frac{1+c_2 \left( \lambda H(a)/k \right)^2}{1+\left( \lambda H(a)/k \right)^2} \right].
\label{SigmaEvolution}
\end{equation}
This time evolution has been used in several recent works including the Dark Energy Survey (DES) \cite{DESMG2018}. The effective dark energy density is used in the ratio such that $\Omega_{DE}(a)/\Omega_\Lambda=1$ today, so the parameters take their GR values today. For the scale dependence, the same reasoning and limits apply as in the previous parametrization. Here and in \cite{DESMG2018}, it is worth noting that GR is recovered when $\mu_0=0$ and $\Sigma_0=0$. Also, in \cite{DESMG2018}, the scale dependence is not modeled and their equations don't have the terms "$1$" as in the RHS of our equations (\ref{muEvolution}) and (\ref{SigmaEvolution}) because they defined their MG equations (\ref{muEquation}) and (\ref{SigmaEquation}) with $1+\mu(a,k)$ and $1+\Sigma(a,k)$ instead. In other words, the two parametrizations are the same except that we absorbed the terms "$1$" from their MG equations into our two parameters. This is handled by \texttt{ISiTGR} and leads to exactly the same constraints on MG parameters (see section \ref{sec:results}).

%%%%%%%%%%%%%%%%%%%%%%%%%%%%%%%%%
\subsubsection{Time and scale functional dependencies for $(Q(a,k),\,\, R(a,k),\,\,D(a,k))$}
We implemented the same time and scale dependencies for these 3 parameters with the same limiting cases and behaviors. The equations read:
\begin{equation}
Q(a,k)=1+Q_{0}\frac{\Omega_{DE}(a)}{\Omega_\Lambda} \left[ \frac{1+c_1 \left( \lambda H(a)/k \right)^2}{1+\left( \lambda H(a)/k \right)^2} \right],
\label{Parametrization:Q}
\end{equation}
\begin{equation}
R(a,k) = 1+R_{0}\frac{\Omega_{DE}(a)}{\Omega_\Lambda} \left[ \frac{1+c_2 \left( \lambda H(a)/k \right)^2}{1+\left( \lambda H(a)/k \right)^2} \right]
\label{Parametrization:R}
\end{equation}
and 
\begin{equation}
D(a,k) = 1+D_{0}\frac{\Omega_{DE}(a)}{\Omega_\Lambda} \left[ \frac{1+c_2 \left( \lambda H(a)/k \right)^2}{1+\left( \lambda H(a)/k \right)^2} \right].
\end{equation}
Again, we can go from the ($Q(a,k)$,$R(a,k)$) to the ($Q(a,k)$,$D(a,k)$) parametrization by using $D(a,k)=Q[1+R(a,k)]/2$, but we recall that we cannot use $\Sigma(a,k)=\mu(a,k)[1+\gamma(a,k)(a,k))]/2$ unless the anisotropic shear stress is zero.  

%%%%%%%%%%%%%%%%%%%%%%%%%%%%%%%%%%%%%%
\subsection{MG parameter binned dependencies in time and scale}

Besides the functional forms described above, \texttt{ISiTGR} implements a binning method where the time (redshift) and scale dependencies of the MG parameters are modeled by different parameters in each bin. 
The advantages of using binning methods have been highlighted in \cite{Dossett2015,ZhaoEtAl2010,Song2011s}.

Moreover, \texttt{ISiTGR} uses two different ways to evolve the MG parameters using  binning methods. The first one is the traditional binning method in which one evolves the MG parameters in two different predefined redshift and scale bins. Unlike some treatments, we provide some extra control in the transition between the scale bins by the addition of hyperbolic functions, see e.g. \cite{Daniel2010MG, Daniel2010MG2,Dossett2015}. The second approach incorporated into \texttt{ISiTGR} is the hybrid method, in which we still have two predefined bins for redshift and scale, but we allow an independent monotonic functional evolution for the MG parameters in each bin, this produces a smoother transition between the bins. In the current version of \texttt{ISiTGR}, we implement binning methods for the ($\mu$,$\gamma$), ($\mu$,$\Sigma$) and ($Q$,$D$) parametrizations. 

\subsubsection{Traditional binning}
Here MG parameters are binned in both redshift, z, and wavenumber (scale), k. A total of four bins are created by using two redshift bins and two scale bins. The scale bins are $k \leq k_c$ and $k > k_c$, while the redshift bins are $0 < z \leq z_{div}$  and $z_{div} < z \leq z_{TGR}$. For redshifts $z > z_{TGR}$ the MG parameters take their GR value of 1 at all scales. 

If $X(a,k)$ represents any MG parameter in \texttt{ISiTGR} (e.g. $\mu$, $\gamma$,  $\Sigma$, $Q$, or $D$), then the binned form of the parameter is written as
\begin{equation}
X(a,k) = \frac{1+X_{z_1}(k)}{2}+\frac{X_{z_2}(k)-X_{z_1}(k)}{2}\tanh\left(\frac{z-z_{div}}{z_{tw}}\right) + \frac{1-X_{z_2}(k)}{2}\tanh\left(\frac{z-z_{TGR}}{z_{tw}}\right),
\end{equation}
where $z_{div}$ is the specific redshift at which the transition between the two bins occurs and $z_{TGR}$ is the redshift below which GR is to be tested. In the \texttt{ISiTGR} code, $z_{TGR}=2z_{div}$ is hard-coded giving equally sized bins. 
Also, $z_{tw}$ acts as a transition width for the hyperbolic tangent function which 
is used in order to make the transition between the bins smooth. It is hard-coded as $z_{tw}=0.05$. The binning in scale is implemented via the two parameters $X_{z_1}(k)$ and $X_{z_2}(k)$ as follows: 
\begin{equation}
X_{z_1}(k) = \frac{X_2+X_1}{2}+\frac{X_2-X_1}{2}\tanh\left(\frac{k-k_c}{k_{tw}}\right)
\end{equation}
and
\begin{equation}
X_{z_2}(k) = \frac{X_4+X_3}{2}+\frac{X_4-X_3}{2}\tanh\left(\frac{k-k_c}{k_{tw}}\right),
\end{equation}
where $k_{tw}$ is the transition width between k bins and is hard-coded as $k_{tw}=k_c /10$. Therefore, since $k_c$ quantifies the scale dependence, such a value for $k_{tw}$  ensures that the transition between bins will occur early, before the scale at which the transition occurs. However, the user can change this feature easily.

The method defines 4 parameters $X_i$ for each parameter in the pairs $(\mu(a,k)$,$\gamma(a,k))$, $(\mu(a,k)$,$\Sigma(a,k))$, $(R(a,k)$,$Q(a,k))$ or $(R(a,k)$,$D(a,k))$, so there are a total of 8 parameters. As we can observe in Table \ref{BinningResults}, most of these can be constrained at almost the same level of significance as for the functional forms.

\subsubsection{Hybrid binning method}
This method keeps the same binning in the redshift as the traditional one described above, but it replaces the binning in scale by a functional form within each redshift bin. In other words, the functions $X_{z_1}(k)$ and $X_{z_2}(k)$ follow a monotonic evolution inside each redshift bin. In this case, the functions $X_{z_1}(k)$ and $X_{z_2}(k)$ are implemented as 
\begin{equation}
X_{z_1}(k) = X_1 e^{-k/k_c} + X_2 (1-e^{-k/k_c})
\end{equation}
and
\begin{equation}
X_{z_2}(k) = X_3 e^{-k/k_c} + X_4 (1-e^{-k/k_c}).
\end{equation}
Of course, the exponential functional form can be replaced by other functions. As shown in the original \texttt{ISiTGR} paper, using hybrid parametrization produces a smoother matter power spectrum.

A summary table of the parametrizations covered by \texttt{ISiTGR} is given in Table VI in Appendix B. An illustrating diagram for the various parametrizations (see Fig. 9), as well as a flowchart that shows the parameter files in \texttt{ISiTGR} (see Fig. 10) are given in Appendix C. 

%%%%%%%%%%%%%%%%%%%%%%%%%%%%%%%%%%%%%%%%%%%%%%%%%%%%%%
\section{Implementation of modified gravity equations in \texttt{ISiTGR} patch to \texttt{CAMB} and \texttt{CosmoMC}}

\label{sec:Implementation}
%%%%%%%%%%%%%%%%%%%%%%%%%%%%%%%%%%%%%%%%%%%%%%%%%%%%%%
%%%%%%%%%%%
%%
\subsection{CMB implementation and the synchronous gauge}

We first describe here the implementation of the MG formalism in the CMB software code \texttt{CAMB} \cite{CAMB} to calculate various CMB temperature anisotropy and polarization auto and cross spectra ($C_\ell^{TT}$, $C_\ell^{TE}$, $C_\ell^{EE}$, $C_\ell^{BB}$) as well as the three-dimensional matter power spectrum $P_\delta(k,z)$. These are powerful probes to  constrain both the growth history of structure in the Universe and the expansion history of the Universe. We describe the overall formalism and some key changes to the code in a self-contained way here but refer the reader for more details in the technical documentation of \texttt{ISiTGR} provided in the github repository \url{https://github.com/mishakb/ISiTGR}. 

We recall that the package \texttt{CAMB} is written in the synchronous gauge where the perturbed FLRW metric is written as:
\be
ds^2=a(\tau)^2[-d\tau^2+(\gamma_{ij}+h_{ij})dx^idx^j],
\label{eq:FLRWSync}
\ee
where $h_{ij}$ represents the metric perturbation in this gauge. The metric potentials are defined from the trace ($h$) and traceless ($h+6\eta$) part of the metric perturbation, following the notation of \cite{Ma-Bertschinger1995}.  $h_{ij}$ is explicitly expanded in terms of $G$, described further above, giving the following form for a single mode \cite{Ma-Bertschinger1995}
\be
h_{ij} = \frac{h}{3}\gamma_{ij}G+(h+6\eta)(k^{-2}G_{|ij}+\frac{1}{3}\gamma_{ij}G).
\ee
Next, combining the perturbed Einstein's equations as discussed to get Eqs. (\ref{eq:Poisson}) and (\ref{eq:2ndEin}), we have in the synchronous gauge \cite{Ma-Bertschinger1995}
\bea
\left(k^2-3K\right)(\eta-\mathcal{H}\alpha)  &=&-4\pi G a^2\sum_i \rho_i \Delta_i,
\label{eq:PoissonSync}\\
k^2(\dot{\alpha}+2\mathcal{H}\alpha-\eta) &=& -12 \pi G a^2\sum_i \rho_i(1+w_i)\sigma_i,
\label{eq:2ndEinSync}
\eea
where $\alpha = (\dot{h}+6\dot{\eta})/2k^2$. Next, using the gauge invariance of $\Delta_i$ and $\sigma_i$ and comparing equations (\ref{eq:Poisson}) and (\ref{eq:2ndEin}) versus (\ref{eq:PoissonSync}) and (\ref{eq:2ndEinSync}), we can see that the potentials in the two gauges are related by

\begin{equation}
\Phi = \eta - \mathcal{H}\alpha
\label{NewSin1}
\end{equation}
and
\begin{equation}
\Psi = \dot{\alpha} + \mathcal{H}\alpha.
\label{NewSin2}
\end{equation} 

\subsection{\texttt{CAMB} variables and potential evolution}
Additionally, \texttt{CAMB} defines two quantities that are used throughout the code when evolving the perturbations. These are 
\begin{equation}
\sigma_{\rm{CAMB}} \equiv k\alpha = \frac{k(\eta-\Phi)}{\mathcal{H}},
\label{sigmacamb}
\end{equation} 
\begin{equation}
\mathcal{Z}_{\rm{CAMB}} \equiv \dot{h}/2k = \sigma_{\rm{CAMB}} -3\frac{\dot{\eta}}{k}
\label{zcamb}
\end{equation} 
and it follows (\ref{NewSin2}) that
\begin{equation}
\dot{\sigma}_{\rm{CAMB}} = k \Psi - \mathcal{H}\sigma_{\rm{CAMB}}.
\label{sigmadot}
\end{equation}

\texttt{CAMB} evolves the metric potential $\eta$ ($k\eta$ exactly) as well as the matter perturbations, $\delta_i$, the heat flux, $q_i$, and
the shear stress $\sigma_i$ (via $\Pi_i=\frac{3}{2}(1+w_i)\sigma_i$) for each matter species in the synchronous gauge according to the evolution equations as described in \cite{Ma-Bertschinger1995} and the documentation of \texttt{CAMB}. Additionally, the \texttt{CAMB} variables, $\sigma_{\rm{CAMB}}$, $\dot{\sigma}_{\rm{CAMB}}$ and $Z_{\rm{CAMB}}$ are evaluated at each time step. Further, we give below the expressions for the key quantities $\dot{\eta}$ and $\dot{\alpha}$ for each pair of MG parameters and taking into account massive neutrino, spatial curvature, and dynamical dark energy equation parameters.  

In this new version of \texttt{ISiTGR} we account consistently for contributions from massive neutrinos and radiation as they enter the terms of the form $8\pi G a^2\sum_i \rho_i\Pi_i (3w_i+1)$ found in the expressions for the time evolution of the potentials, where the anisotropic shear $\Pi_i$ is defined as $\Pi_i = \frac{3}{2}(1+w_i)\sigma_i$.

 \subsubsection{$\dot{\eta}$ for various MG parametrizations}

Taking into account the specific modifications to the growth equations in each parametrization and not neglecting the anisotropic shear contribution leads to different expressions for  $\dot{\eta}$. We note again that some relationships like (\ref{eq:Sigma2}) only hold when the anisotropic shear contribution is  neglected. We provide the full expressions below.  

First, taking the time derivative of  (\ref{NewSin1}) gives
\begin{equation}
\dot{\eta} = \dot{\mathcal{H}}\alpha + \mathcal{H}\dot{\alpha} + \dot{\Phi}.
\label{etadot1}
\end{equation}
Next, we need to analyze each parametrization separately. 

For the $(\mu,\gamma)$ pair, we take the derivative of (\ref{gammaEquation}) and put it into (\ref{etadot1}) to obtain
\begin{equation}
\begin{split}
\dot{\eta}_{(\mu,\gamma)} & = \dot{\mathcal{H}}\alpha + \mathcal{H}\dot{\alpha} + \frac{\dot{\mu}}{2k^2}\sum_i \bigg[ 2\Pi_i\hat{\rho_i}(1-\gamma) - \gamma\beta_k \Delta_i \hat{\rho_i} \bigg] +  \frac{\mu}{2k^2}\sum_i \bigg\{ 2\dot{\Pi_i}\hat{\rho_i}(1-\gamma) + \\
& \dot{\hat{\rho_i}}\left[ 2\Pi_i(1-\gamma)-\gamma\beta_k \Delta_i\right] - \dot{\gamma}\hat{\rho_i}(\beta_k \Delta_i+2\Pi_i)-\gamma\beta_k \hat{\rho_i}\dot{\Delta_i} \bigg\},
\end{split}
\label{etadot2}
\end{equation}
where we have defined $\beta_k \equiv (1-\frac{3K}{k^2})^{-1}$ with $\beta_k=1$ for the spatially flat case. We have also added the subscript label $(\mu$,$\gamma$) to specify that we are deriving the expression for $\dot{\eta}$ in this particular parametrization. Using (\ref{NewSin2}) and (\ref{muEquation}), we get
\begin{equation}
\dot{\alpha} = - \frac{\mu}{2k^2}\sum_i\bigg[\beta_k \Delta_i\hat{\rho_i} + 2\Pi_i\hat{\rho_i}\bigg] - \mathcal{H}\alpha.
\label{alphadot}
\end{equation}
In order to work with the quantities used by \texttt{CAMB}, we rewrite (\ref{eq:Deltadot}) as
\begin{equation}
\dot{\Delta_i} = 3(1+w_i)(\dot{\Phi}+\mathcal{H}\Psi) + 3\mathcal{H}w_i\Delta_i - 2\mathcal{H}\Pi_i - kq_i^{(N)}f_1,
\label{Deltadot}
\end{equation}
where $f_1 \equiv 1+\frac{3(\mathcal{H}^2-\dot{\mathcal{H}})}{k^2}$ and, for now, we use the superscript $(N)$ to denote that in this relation the heat flux $q_i$ is still given in the Newtonian gauge; this will be dealt with below. Note that we now use $\dot{\Delta}_i$ for the  anisotropic stress used in \texttt{CAMB}.

Next, putting equations (\ref{alphadot}) and (\ref{Deltadot}) into (\ref{etadot2}), and 
 using the continuity equation
 \begin{equation}
 \dot{\hat{\rho_i}} = -\mathcal{H}\hat{\rho_i}(1+3w_i),
 \label{rhodot}
 \end{equation} as well as Eqs. (\ref{NewSin1}) and (\ref{NewSin2}), and the relationship
 \begin{equation}
 q_i^{(N)} = q_i^{(S)} + k\alpha(1+w_i),
 \label{heatflux}
 \end{equation}
 for $q_i$ between the two gauges, we obtain the final expression for $\dot{\eta}_{(\mu,\gamma)}$ as
\begin{equation}
\begin{split}
\dot{\eta}_{(\mu,\gamma)} & = \frac{1}{2f_{\mu,\gamma}} \bigg\{ k\mu\gamma\beta_k f_1 \sum_i q_i^{(S)}\hat{\rho_i} + \sum_i \beta_k \Delta_i\hat{\rho_i}[\mathcal{H}\mu(\gamma-1)-\dot{\mu}\gamma - \mu\dot{\gamma}] \\& 
 + 2\mu(1-\gamma)\sum_i\dot{\Pi_i}\hat{\rho_i} + k^2\alpha [-2(\mathcal{H}^2-\dot{\mathcal{H}})+\mu\gamma\beta_k \sum_i\hat{\rho_i}(1+w_i)] \\&
 - 2[\mu\dot{\gamma}+\dot{\mu}(\gamma-1)]\sum_i\Pi_i\hat{\rho_i} - 2\mathcal{H}\mu\sum_i\Pi_i\hat{\rho_i}(3w_i+2) \\&
 +2\mathcal{H}\mu\gamma\sum_i \hat{\rho_i}\Pi_i (3w_i+1+\beta_k)
\bigg\},
\end{split}
\label{etadot5}
\end{equation}
where we defined 
\begin{equation}
f_{\mu,\gamma} \equiv k^2+\frac{3}{2}\beta_k  \mu\gamma\sum_i(1+w_i)\hat{\rho_i}. 
\end{equation}

Following similar steps, using the corresponding equations for the $(\mu,\Sigma)$ parametrization, we derive
\begin{equation}
\begin{split}
\dot{\eta}_{(\mu,\Sigma)} & = \frac{1}{2f_{\mu,\Sigma}} \bigg\{ k\beta_k(2\Sigma-\mu) f_1 \sum_i q_i^{(S)}\hat{\rho_i} + \beta_k[(\dot{\mu}-2\dot{\Sigma})+2\mathcal{H}(\Sigma-\mu)]\sum_i \Delta_i\hat{\rho_i} \\&
+ 2(\mu-\Sigma)\sum_i\dot{\Pi_i}\hat{\rho_i}  + 2[(\dot{\mu}-\dot{\Sigma})+\mathcal{H}\beta_k(2\Sigma-\mu) -\mathcal{H}\mu]\sum_i\Pi_i\hat{\rho_i} \\&
+ 2\mathcal{H}(\Sigma-\mu)\sum_i\Pi_i(1+3w_i)\hat{\rho_i} + k^2\alpha \left[ \beta_k(2\Sigma-\mu)\sum_i(1+w_i)\hat{\rho_i}-2(\mathcal{H}^2-\dot{\mathcal{H}}) \right] 
\bigg\},
\end{split}
\label{etadot7}
\end{equation}
where 
\be 
f_{\mu,\Sigma}\equiv k^2+\frac{3}{2}\beta_k(2\Sigma-\mu)\sum_i\hat{\rho_i}(1+w_i).
\ee

Again, it is important to recall that we cannot go from (\ref{etadot7}) to (\ref{etadot5}) just by using the usual relation $\Sigma=\frac{\mu}{2}(1+\gamma)$, because this is only valid when contributions to the anisotropic stress are zero for all species. These are more general equations including non-zero shear terms, and they are valid for dynamical dark energy and for flat or curved spaces.

Finally, we derive the expression for the $(Q(a,k),D(a,k))$ parametrization as
\begin{equation}
\begin{split}
\dot{\eta}_{(Q,D)} & = -\frac{1}{2f_Q} \bigg\{ 2k^2\alpha(\mathcal{H}^2-\dot{\mathcal{H}}) + [2 \mathcal{H}(D-Q)+\dot{Q}]\beta_k \sum_i \Delta_i \hat{\rho}_i \\&
- k^2\alpha \sum_i \beta_k Q(1+w_i)\hat{\rho}_i -k\beta_k Qf_1\sum_i q_i^{(S)}\hat{\rho}_i - 2\mathcal{H}Q(\beta_k-1)\sum_i \hat{\rho_i}\Pi_i
\bigg\},
\end{split}
\label{etadot8}
\end{equation}
where 
\be
f_Q\equiv k^2+\frac{3}{2}\beta_k Q\sum_i(1+w_i)\hat{\rho}_i.
\ee
In this case, this expression is also valid for the pair $(Q(a,k),R(a,k))$ upon using that $R(a.k)=\frac{2D(a,k)}{Q(a,k)}-1$, which is a definition rather than a relationship.

Finally, we made some comparisons of the angular and matter power spectra ($C_{\ell}^{TT}$ and $P(k)$) between \texttt{ISiTGR} and  \texttt{MGCAMB}  \cite{MGCAMB3} and found them to be in an agreement to $10^{-4}$ or better for some of the scale-independent parametrizations listed in section  \ref{sec:MGP}.

%%%%%%%%%%%%%%%%%%%%%%%%%%%%%%%%%%%%%%%%%%%%%%%%%%%%%%%%%

{
\section{Modified power spectra and correlation functions}

\subsection{$3 \times 2$ point power spectra and correlation functions}

%%%%%%%%%%%%%%%%%%%%%%%%%%%%%%%%%%%%%%%%%%%%%%%%%%%%%%%%%
The current version of \texttt{ISiTGR} consistently implements the often-used $3 \times 2$ point statistics (lensing -- lensing, galaxy -- galaxy lensing and galaxy -- galaxy) by modifying the respective transfer functions via modification to the Weyl potential (for the first two) satisfying, for example, the MG equation (\ref{SigmaEquation}). The angular power spectra are then calculated as 
\begin{equation}
P^{ij}_{\kappa\kappa}(\ell) = \int_0^{\chi_H} d\chi \frac{q^i_\kappa (\chi) q^j_\kappa (\chi)}{\chi^2} P_{\Phi_{W},\Phi_{W}}\left(\frac{\ell+1/2}{f_K(\chi)}, \chi \right),
\end{equation}
for lensing -- lensing and  
\begin{equation}
P^{ij}_{\delta\kappa}(\ell) = \int_0^{\chi_H} d\chi \frac{q^i_\delta \left(\frac{\ell+1/2}{f_K(\chi)},\chi\right) q^j_\kappa (\chi)}{\chi^2} P_{\delta, {\phi_W}}\left(\frac{\ell+1/2}{f_K(\chi)}, \chi \right),
\end{equation}
for the galaxy--galaxy lensing, where 
$P_{\Phi_{W},\Phi_{W}}$ is the Weyl potential power spectrum, and 
$f_K(\chi)$ here is the comoving angular diameter distance. The lensing efficiency function  and the radial weight function are given by
\begin{equation}
q^i_\kappa(\chi) = \chi \int_\chi^{\chi_H} d\chi' n^i(\chi) \frac{f_K(\chi' - \chi)}{f_K(\chi)}
\end{equation}
and 
\begin{equation}
q^i_\delta \left(\frac{\ell+1/2}{f_K(\chi)},\chi\right) = b^i(k,z(\chi)) n^i(\chi), 
\end{equation}
respectively. For the radial weight function, $b^i(k,z(\chi))$ corresponds to the galaxy bias. It is worth mentioning that we are using  the two Newtonian Gauge potentials directly to compute the power spectrum, so we do not use the equality $k^2\Phi_W = -\frac{3}{2}(1+z)\left(\frac{H_0}{c}\right)\Omega_m\delta$, which is only valid when $\Phi=\Psi$.

The corresponding lensing--lensing (or shear-shear) 2-point correlation functions are calculated as 
\begin{equation}
\xi_{\pm}^{ij}(\theta) = \int \frac{d\ell\,\, \ell}{2\pi} P^{ij}_{\kappa\kappa}(\ell) J_{2\pm2} (\ell\theta),
\end{equation}
where $J_n$ is the $n^{th}$ order Bessel function of the first kind. Similarly, the galaxy -- galaxy lensing 2-point correlation function is given by 
\begin{equation}
\gamma_t^{ij}(\theta) = \int \frac{d\ell\,\, \ell}{2\pi} P^{ij}_{\delta\kappa}(\ell) J_2 (\ell\theta)
\end{equation}
and the clustering 2D correlation function is given by 
\begin{equation}
w^i(\theta) = \int \frac{d\ell\,\, \ell}{2\pi} P^{ii}_{\delta\delta}(\ell) J_0 (\ell\theta),
\end{equation}
where $P^{ii}_{\delta\delta}(\ell)$ is the matter power spectrum. We refer the reader to, for example, \cite{Krause2016,DES2017} for details on  the $3 \times 2$ point formalism. 
}

%%%%%%%%%%%%%%%%%%%%%%%%%%%%%%%%%%%%%%%%%%%%%%%%%%%%%%%%%
\subsection{Integrated Sachs-Wolfe (ISW) effect}

{
In order to propagate the MG equation changes to the ISW effect, we follow \texttt{CAMB}'s implementation to calculate the derivatives of MG potentials. This is done for $\dot{\Phi}$ by recalling Eq. (\ref{etadot1}) and using $\dot{\eta}$ and $\alpha=\frac{\sigma_{\rm{CAMB}}}{k}$ as
\begin{equation}
\dot{\Phi} = \dot{\eta} - \dot{\mathcal{H}}\alpha - \mathcal{H} \dot{\alpha}.
\label{Phidot1}
\end{equation}
Then, one obtains $\dot{\Psi}$ by directly taking the derivative of (\ref{muEquation}) which depends only on $\mu$, so it can be used for both the $(\mu,\gamma)$ and $(\mu,\Sigma)$ parametrizations. This gives
\begin{equation}
\dot{\Psi} = -\frac{\dot{\mu}}{2k^2}\sum_i \left[\beta_k\Delta_i\hat{\rho_i} + 2\Pi_i\hat{\rho_i}\right] + \frac{\mu}{2k^2}\sum_i\left[(\beta_k\Delta_i+2\Pi_i)\dot{\hat{\rho_i}} + (\dot{\beta_k\Delta_i}+2\dot{\Pi_i})\hat{\rho_i} \right].
\end{equation}
Next, we can substitute (\ref{Deltadot}), (\ref{rhodot}) and (\ref{heatflux}) into the above equation to obtain
\begin{equation}
\begin{split}
\dot{\Psi} & = -\frac{\dot{\mu}}{2k^2}\sum_i \left[\beta_k\Delta_i\hat{\rho_i} + 2\Pi_i\hat{\rho_i}\right] + \frac{\mu}{2k^2}\sum_i \bigg\{ \mathcal{H}\beta_k\Delta_i\hat{\rho_i} - 2\dot{\Pi_i}\hat{\rho_i} + k \beta_k f_1 q_i^{(S)} \hat{\rho_i} \\ &
+\beta_k(1+w_i)\hat{\rho_i}[k^2\alpha f_1-3(\dot{\Phi}+\mathcal{H}\Psi)] + 2\mathcal{H}\Pi_i\hat{\rho_i}(1+3w_i) + 2\beta_k\mathcal{H}\hat{\rho_i}\Pi_i \bigg\}.
\end{split}
\label{Psidot2}
\end{equation}
Thus, the MG changes to the ISW are propagated through (\ref{Phidot1}) and (\ref{Psidot2}). 

Finally, for the ISW effect for the $(Q,D)$ and $(Q,R)$ parametrizations, we derive and use another expression for $\dot{\Psi}$. Following similar steps we can obtain
						
\begin{equation}
\begin{split}
\dot{\Psi} & = - \dot{\Phi} - \frac{1}{k^2}\sum_i\bigg\{
\beta_k\dot{D}\Delta_i\hat{\rho}_i+\dot{Q}\Pi_i\hat{\rho}_i - \beta_k D\mathcal{H}\Delta_i\hat{\rho}_i - Q\mathcal{H}\Pi_i(3w_i+1)\hat{\rho_i} + Q\dot{\Pi}_i\hat{\rho}_i \\&
- 2\beta_k D\mathcal{H}\Pi_i\hat{\rho}_i- k\beta_k Df_1 q_i^{(S)}\hat{\rho}_i + \beta_k(1+w_i)\hat{\rho_i}[3D(\dot{\Phi}+\mathcal{H}\Psi) - k^2\alpha f_1 D]
\bigg\}.
\end{split}
\label{Psidot3}
\end{equation}
Here, it is important to mention that firstly, the code computes $\dot{\Phi}$ and afterwards, it computes $\dot{\Psi}$ in terms of $\dot{\Phi}$.
%%%%%%%%%%%%%%%%%%%%%%%%%%%%%%%%%%%%%
}
%%%%%%%%%%%%%%%%%%%%%%%%%%%%%%%%%%%%%
%

%%%%%%%%%%%%%%%%%%%%%%
\section{Applications and results from current available data sets}%%%%%
%%%%%%%%%%%%%%%%%%%%%%

\label{sec:results}

{In this section we apply \texttt{ISiTGR} to current cosmological data sets using different MG parametrizations. As a start, we reproduce and compare to some of the results from the Planck-2018 \cite{Planck2018}, Planck-2015 \cite{Planck2015MG} and DES-2018 \cite{DESMG2018} papers about extended models. We then derive new results involving the different features of \texttt{ISiTGR} such as spatial curvature, dynamical dark energy, and massive neutrinos along with MG parameters. Moreover, we derive constraints and correlations for the new binning methods implemented in \texttt{ISiTGR} for MG parameters.

For all results, in addition to MG parameters, we also vary the six core cosmological parameters: $\Omega_bh^2$ and $\Omega_c h^2$, the baryon and cold dark matter physical density parameters, respectively; $\theta$, the ratio of the sound horizon to the angular diameter distance of the surface of last scattering; $\tau$, the reionization optical depth; $n_s$, the spectral index; and $\ln(10^{10} A_s)$, the amplitude of the primordial power spectrum.

%We use three different parametrizations of the MG parameters:  two scale-dependent methods including a traditional binning method and a hybrid evolution method, both of which were discussed in our previous work \cite{r40}, and a scale independent method using a functional first introduced by \cite{r33}:  

\subsection{Data sets}
We combine several current data sets including measurements from the Cosmic Microwave Background (CMB) Planck mission. Specifically, the Planck-2018 TT, TE and EE spectra at $\ell \geq 30$ likelihood, the low-$\ell$ temperature Commander likelihood, the low-$\ell$ SimAll EE likelihood and CMB lensing measurements, used in \cite{Planck2018}. We label this combination as TTTEEE + lowE + CMBlens (2018). Moreover, we use the Planck-2015 likelihood presented in \cite{Planck2015}, where the Planck TT likelihood for multipoles $30 \leq \ell \leq 2508$ and the joint TT, EE, BB, and TE likelihood for $2 \leq \ell \leq 30$ were used. We refer to this likelihood combination as TT + lowP (2015)$^2$. \footnotetext[2]{We use the public Planck likelihood package files plik\_dx11dr2\_HM\_v18\_TT.clik and lowl\_SMW\_70\_dx11d\_2014\_10\_03\_v5c\_Ap.clik, unless otherwise specified.}We also use in some cases the Planck-2015 CMB lensing measurements for temperature from \cite{Planck2015XV}.
We also use the likelihood included into \texttt{CosmoMC} for clustering and lensing data from the Dark Energy Survey (DES) Year 1 \cite{DES2017}, but with the non-linear data points removed since these are not yet accurately modeled for MG theories, see e.g. \cite{DESMG2018} for a discussion. We also add measurements coming from Type Ia supernovae. Namely, the Pantheon sample data presented in \cite{Pan-STARRS-2017}, which combines 279 SNe Ia $(0.03< z <0.68)$ with useful distance estimates of SNe Ia from SDSS, SNLS, various low-z and HST samples, giving a total of 1048 SNe Ia ranging from $(0.01< z <2.3)$. In this work, we refer to these data sets as DES and Pantheon, respectively. Finally, we also consider measurements from Baryon Acoustic Oscillations (BAO) and Redshift Space Distortions (BAO/RSD). Specifically, we use BAO and BAO/RSD measurements coming from the BOSS Data Release 12 \cite{AlamEtAl2016}. Moreover, we use another two BAO data sets from the 6DF Galaxy Survey \cite{BeutlerEtAl2011} and the SDSS Data Release 7 Main Galaxy Sample \cite{RossEtAl2014}.

\subsection{Constraints and correlations}
In the case of the $(\mu,\gamma)$ parametrization, we reproduce some results for the constraints and scale dependence as shown in Planck's 2015 dark energy and modified gravity paper \cite{Planck2015MG}, while for the $(\mu,\Sigma)$ parametrization we reproduce some of the results of the DES 2018 constraints on extended cosmological models. Then, we add the new features of \texttt{ISiTGR} in combination with the MG parameters. Finally, we obtain constraints and correlations for the binning methods for $(\mu,\gamma)$ and $(\mu,\Sigma)$.

\subsubsection{Results for $(\mu,\gamma)$ parametrization}
We evolve these parameters in time and scale according to Eqs. (\ref{muEvolutionPlanck}) and (\ref{gammaEvolutionPlanck}) as in, for example, \cite{Planck2015MG}. Our results are shown in Fig. \ref{Plots_Planck2015} and Table \ref{TableResults1}, and are in good agreement for the time dependence with \cite{Planck2018} and \cite{Planck2015MG} but we find some differences when adding scale dependence, see Fig. \ref{Plot_scaledependence}. Specifically, the contour plots agree for the large scales with $k=10^{-7}$Mpc$^{-1}$ but not for small scales with large $k=10^2$Mpc$^{-1}$ as shown in their figure 18 \cite{Planck2015MG}. With a closer look, one can see that for such large $k$ the MG parameters should reduce to 
 $\mu \rightarrow 1+E_{11}\Omega_{DE}$ and $\gamma \rightarrow 1+E_{22}\Omega_{DE}$, so they becomes similar to the scale independent case, as we find in our Fig. \ref{Plot_scaledependence}.
 For large scales (small $k$) the MG parameters should reduce to 
 $\mu \rightarrow 1+c_1 E_{11}\Omega_{DE}$ and $\gamma \rightarrow 1+c_2 E_{22}\Omega_{DE}$. Therefore, after marginalizing over $c_1$ and $c_2$ constraints become weaker and our results are in agreement for this regime. This difference and further study of scale-dependence will be explored further in a separate study.  

%\begin{figure}[t]
%\begin{center}
%\begin{tabular}{ c c }
% {\includegraphics[width=8.8cm]{Figures/Planck2015_1_updated.pdf}} 
% {\includegraphics[width=8.8cm]{Figures/Planck2015_2_updated.pdf}}
%\end{tabular}
%\end{center}
%\caption{Left: 68\% and 95\% confidence contour plots for $\mu_0$ and $\eta_0$ in the $(\mu,\gamma)$ parametrization with DE-time evolution and no scale dependence. This is in good agreement with results from Planck-2015 \cite{Planck2015MG} (see left panel of Fig. 14 there). Right: Contour plots for $\mu_0$ and $\Sigma_0$ in the $(\mu,\Sigma)$ parametrization with similar evolution. This is again in good agreement with Fig. 15 of Planck-2015 \cite{Planck2015MG}}
%\label{Plots_Planck2015}
%\end{figure}

%\begin{figure}[t]
%\begin{table}[h!]
%\begin{center}
% \begin{tabular} { c c }
%  {\includegraphics[width=4cm]{Figures/Planck2015_1_updated.pdf}} 
% &  {\includegraphics[width=4cm]{Figures/Planck2015_2_updated.pdf}} \\
%  {\includegraphics[width=4cm]{Figures/Planck2018_1_updated.pdf}} 
% &  {\includegraphics[width=4cm]{Figures/Planck2018_2_updated.pdf}} \\
%\end{tabular}
%\end{figure}
%\end{center}
%\caption{Left: 68\% and 95\% confidence contour plots for $\mu_0$ and $\eta_0$ in the $(\mu,\gamma)$ parametrization with DE-time evolution and no scale dependence. This is in good agreement with results from Planck-2015 \cite{Planck2015MG} (see left panel of Fig. 14 there). Right: Contour plots for $\mu_0$ and $\Sigma_0$ in the $(\mu,\Sigma)$ parametrization with similar evolution. This is again in good agreement with Fig. 15 of Planck-2015 \cite{Planck2015MG}}
%\end{table}

\begin{figure}[h!]
\begin{tabular}{ c c }
  {\includegraphics[width=7.5cm]{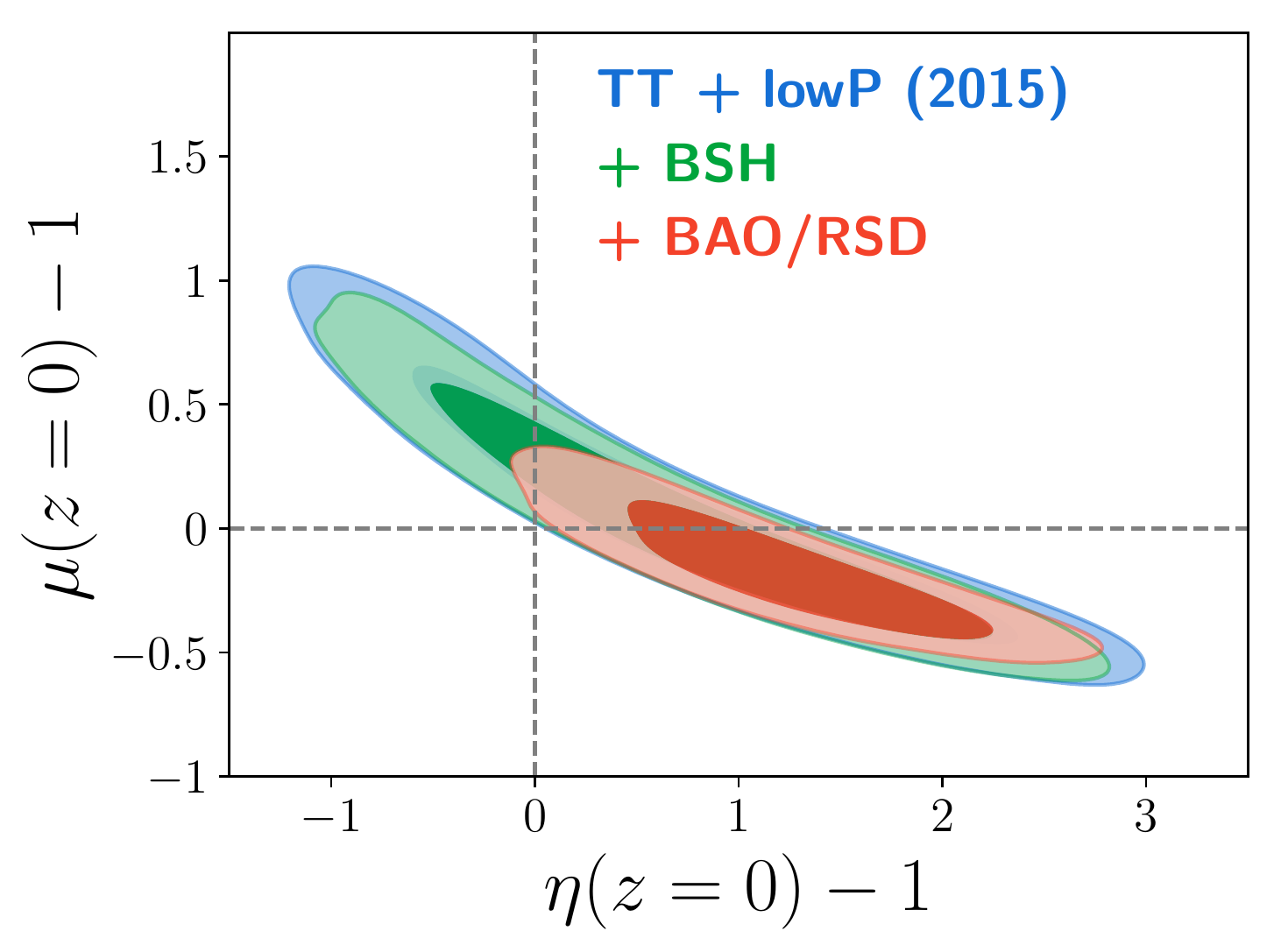}} 
 &  {\includegraphics[width=7.5cm]{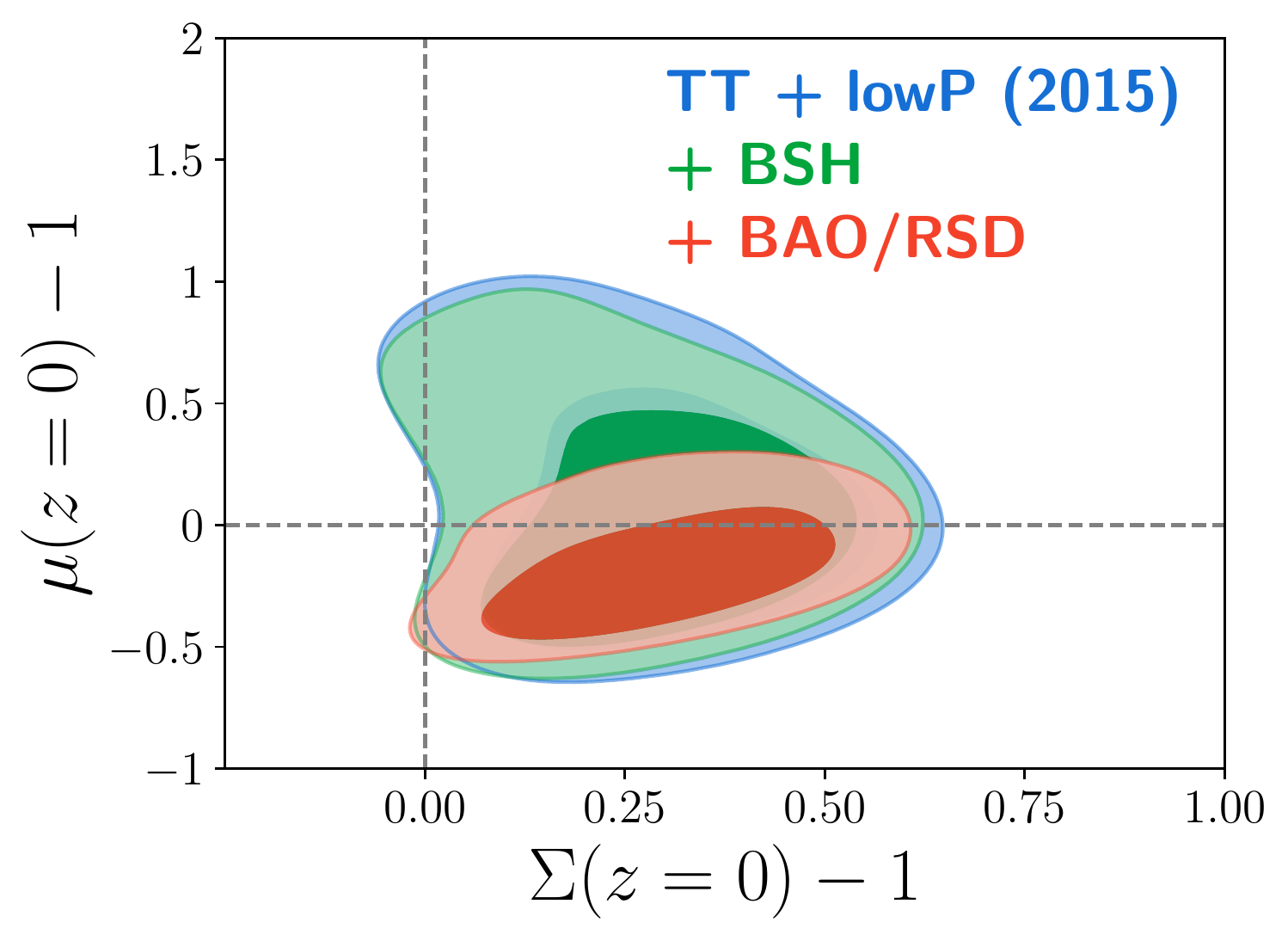}} \\
  {\includegraphics[width=7.5cm]{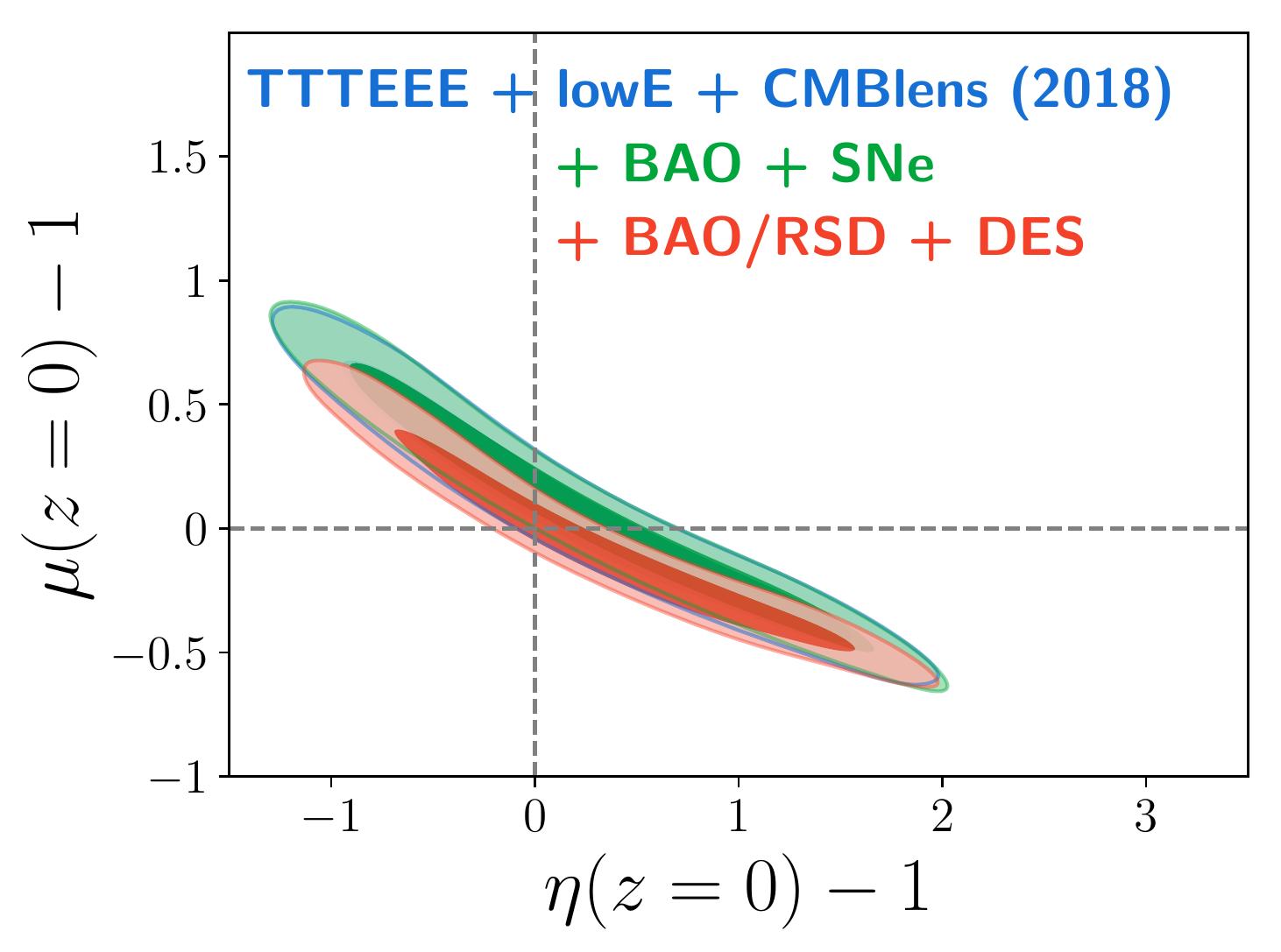}} 
 &  {\includegraphics[width=7.5cm]{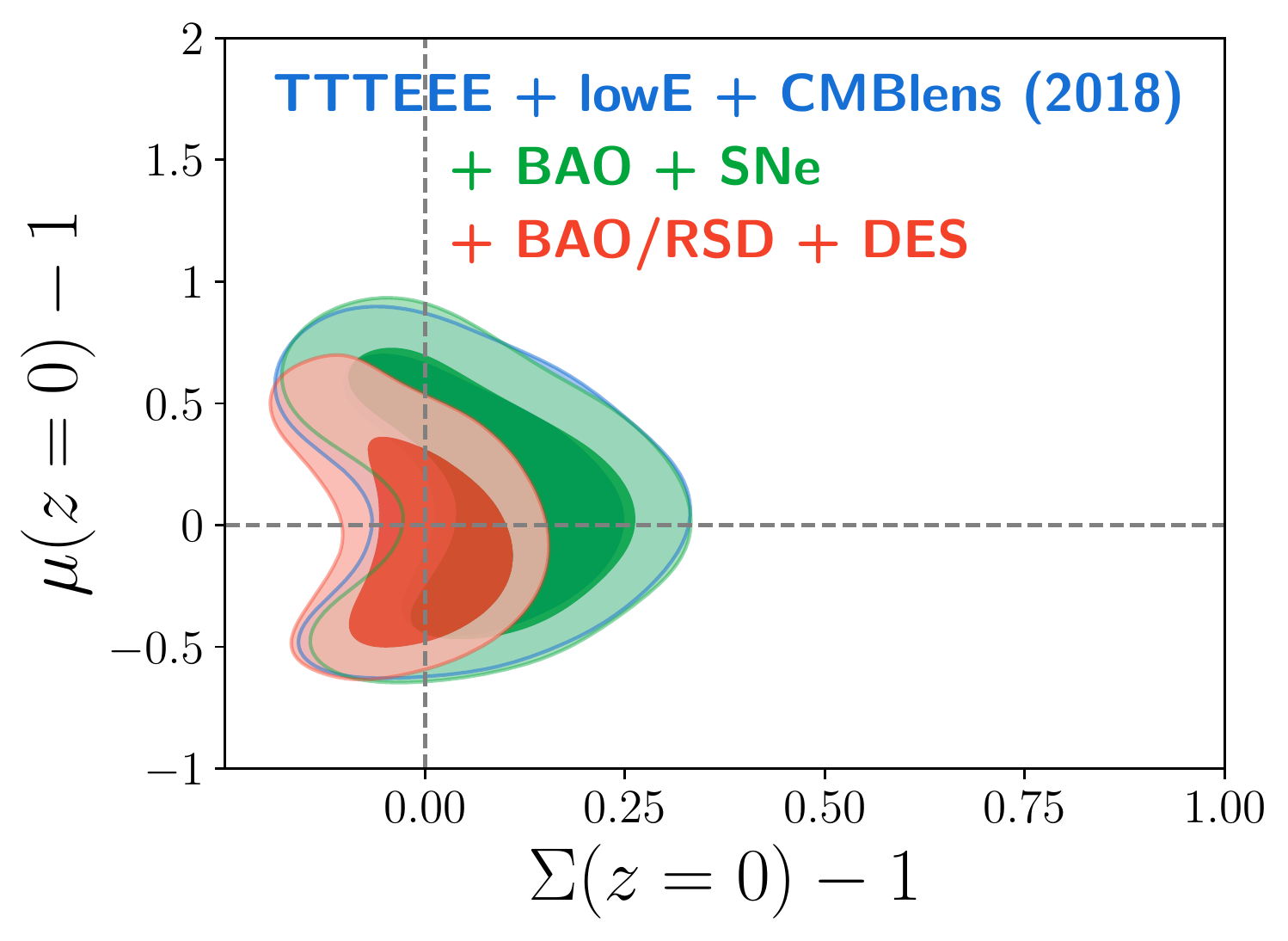}} \\
\end{tabular}
\caption{Left: 68\% and 95\% confidence contour plots for $\mu_0$ and $\eta_0$ in the $(\mu,\gamma)$ parametrization with DE-time evolution and no scale dependence. Figures at the top show constraints with Planck-2015 data while figures at the bottom show constraints using Planck-2018 data. This is in good agreement with results from Planck-2015 \cite{Planck2015MG} (see left panel of Fig. 14 there) and Planck-2018 \cite{Planck2018}. Right: Contour plots for $\mu_0$ and $\Sigma_0$ in the $(\mu,\Sigma)$ parametrization with similar evolution. This is again in good agreement with Fig. 15 of Planck-2015 \cite{Planck2015MG} and Planck-2018 \cite{Planck2018} (see our Table \ref{TableResults1}).}
\label{Plots_Planck2015}
\end{figure}

%%%%%%%%%%%%%%%%
\begin{table}[t!]
\begin{center}
\scriptsize 
 \begin{tabular} { p{1.8cm}  | >{\centering}m{2.5cm} >{\centering}m{2.5cm} >{\centering}m{2.5cm} >{\centering}m{2.5cm} >{\centering}m{2.5cm} >{\raggedleft\arraybackslash}m{2.6cm}}
\hline\hline \\ [-7pt]
 Parameter &  Planck-2015  &  Planck-2015 + BSH  &  Planck-2015 + BAO/RSD & Planck-2018 & Planck-2018 + BAO + SNe & Planck-2018 + BAO/RSD + DES \\[5pt]
\hline
 & & & & & & \\[-7pt]
$E_{11}$ & $0.07^{+0.31}_{-0.69}$ & $0.03^{+0.30}_{-0.63}$ & $-0.25^{+0.19}_{-0.32}$ & $0.13^{+0.46}_{-0.66}$ & $0.12^{+0.42}_{-0.71}$ & $-0.10^{+0.29}_{-0.53}$ \\[5pt]

$E_{22}$ & $1.1\pm 1.4$ & $1.1\pm 1.3$ & $1.80\pm 0.88$ & $0.30^{+0.77}_{-1.6}$ & $0.4\pm 1.1$ & $0.5\pm 1.0$ \\[5pt]

$\mu(z=0)-1$ & $0.05^{+0.22}_{-0.48}$ & $0.02^{+0.21}_{-0.44}$ & $-0.18^{+0.14}_{-0.22}$ & $0.09^{+0.31}_{-0.46}$ & $0.09^{+0.29}_{-0.50}$ & $-0.07^{+0.20}_{-0.36}$ \\[5pt]

$\eta(z=0)-1$ & $0.77^{+0.94}_{-1.1}$ & $0.78\pm 0.88$ & $1.25\pm 0.62$ & $0.20^{+0.50}_{-1.1}$ & $0.25\pm 0.80$ & $0.35\pm 0.70$ \\[5pt]

$\Sigma(z=0)-1$ & $0.29\pm 0.15$ & $0.28\pm 0.14$ & $0.29\pm 0.13$ & $0.07^{+0.12}_{-0.10}$ & $0.082^{+0.12}_{-0.096}$ & $-0.007^{+0.076}_{-0.060}$ \\[5pt]

{$\Omega_b h^2$} & {$0.02253\pm 0.00027$} & {$0.02253 \pm 0.00021$} & {$0.02246 \pm 0.00022$} & {$0.02244 \pm 0.00015$} & {$0.02249 \pm 0.00014$} & {$0.02247 \pm 0.00014$} \\[5pt]

{$\Omega_c h^2$} &{$0.1173\pm 0.0025$} & {$0.1176\pm 0.0012$} & {$0.1179\pm 0.0014$} & {$0.1190\pm 0.0015$} & {$0.1185\pm 0.0010$} & {$0.1187\pm 0.0010$} \\[5pt]

{$100\theta$} & {$1.04126\pm 0.00051$} & {$1.04121\pm 0.00044$} & {$1.04115\pm 0.00044$} & {$1.04104\pm 0.00031$} & {$1.04110\pm 0.00029$} & {$1.04103\pm 0.00029$} \\[5pt]

${\rm{ln}}(10^{10} A_s)$ & $3.057\pm 0.040$ & $3.058\pm 0.041$ & $3.052\pm 0.038$ & $3.031^{+0.018}_{-0.016}$ & $3.032\pm 0.017$ & $3.042^{+0.015}_{-0.017}$ \\[5pt]

$n_s$ & $0.9719\pm 0.0072$ & $0.9711\pm 0.0046$ & $0.9698\pm 0.0047$ & $0.9665\pm 0.0047$ & $0.9679\pm 0.0040$ & $0.9669\pm 0.0039$ \\[5pt]

$\tau$ & $0.064\pm 0.020$ & $0.064\pm 0.021            $ & $0.061\pm 0.019$ & $0.0494^{+0.0083}_{-0.0072}$ & $0.0504\pm 0.0079$ & $0.0548^{+0.0073}_{-0.0082}$ \\[5pt]

$H_0                       $ & $68.6\pm 1.2$ & $68.44\pm 0.57             $ &  $68.26\pm 0.66$ & $67.80\pm 0.66$ & $68.04\pm 0.46$ & $67.93\pm 0.46$ \\[5pt]

$\sigma_8                  $ & $0.816^{+0.033}_{-0.052}$ & $0.814^{+0.029}_{-0.048}   $ & $0.792^{+0.020}_{-0.024}$ & $0.813^{+0.031}_{-0.046}$ & $0.811^{+0.028}_{-0.050}$ & $0.800^{+0.023}_{-0.036}$ \\[5pt]
\hline\hline
\end{tabular}
\end{center}
\caption{Marginalized mean values and 1-$\sigma$ errors for cosmological parameters from various data set combinations. We refer to the combination TTTEEE + lowE + CMBlens (2018) as Planck-2018, while we label TT + lowP (2015) as Planck-2015.}
\label{TableResults1}
\end{table}

For the $(\mu,\gamma)$ parametrization without scale dependence, we set $\lambda=0$ and vary the parameters $E_{11}$ and $E_{22}$ from which $\mu$ and $\gamma$ are constructed. We do not include the weak lensing plots here since the weak lensing likelihood has been replaced by the DES likelihood implemented in the current version of \texttt{CosmoMC}. Also, Planck-2018 reported a bug in the Weyl potential which may  have affected the previous weak lensing results in \cite{Planck2015MG}. It is worth mentioning that, for this specific case, we use the combination of BAO from \cite{AndersonEtAl2014}, Supernovae Type Ia data from the Joint Light-Curve Analysis \cite{BetouleEtAl2013}, and the Hubble constant measurements obtained from the Hubble Space Telescope in \cite{RiessEtAl2011}. We refer to this combination as BSH, as in previous analysis \cite{Planck2015MG}.

\begin{figure}[t]
\begin{center}
 {\includegraphics[width=12cm]{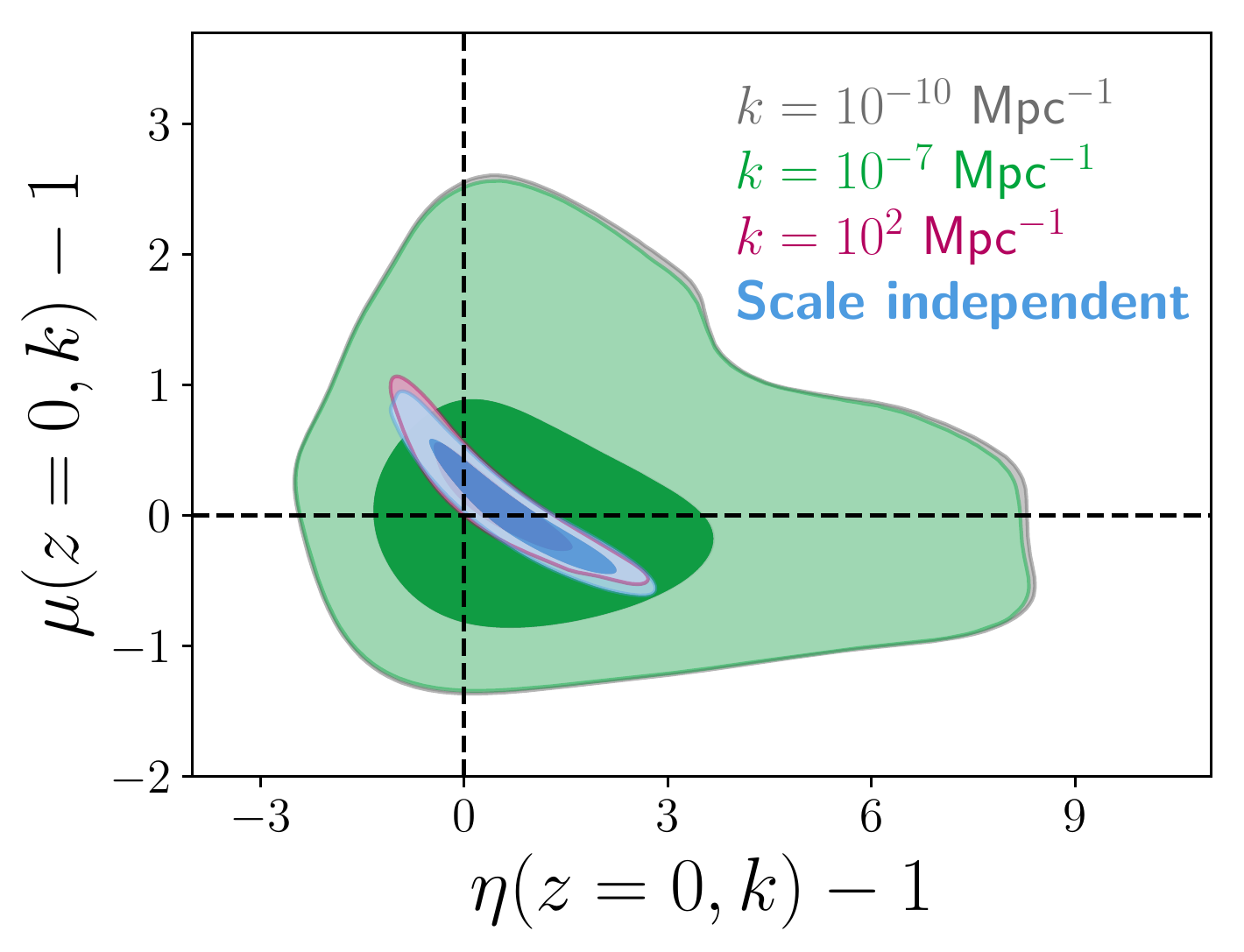}} 
\end{center}
\caption{Results on scale- dependent evolution for $\mu$ and $\gamma$ (called $\eta$ in \texttt{ISiTGR}). We find that for large scales the constraints become weaker than in the scale-independent case, as expected. Here, we plot the cases in which $k=10^{-7}$ Mpc$^{-1}$ and $k=10^{-10}$ Mpc$^{-1}$, as well as the scale-independent case , all for $z=0$. As we mentioned in section \ref{subsec:MGparamz}, for large scales $\mu \rightarrow 1+c_1 E_{11}\Omega_{DE}$ and $\gamma \rightarrow 1+c_2 E_{22}\Omega_{DE}$. Therefore, after marginalizing over $c_1$ and $c_2$ constraints become weaker. However, for small scales we have that $\mu \rightarrow 1+E_{11}\Omega_{DE}$ and $\gamma \rightarrow 1+E_{22}\Omega_{DE}$, then practically this reduce to the scale independent case as we found. The plot in this case of $k=10^{2}$ Mpc$^{-1}$ is not in agreement with that of figure 18 of Planck-2015 \cite{Planck2015MG}. See text for discussion.}
\label{Plot_scaledependence}
\end{figure}

\begin{figure}[t!]
\begin{center}
\begin{tabular}{ c c c }
 {\includegraphics[width = 5.8 cm]{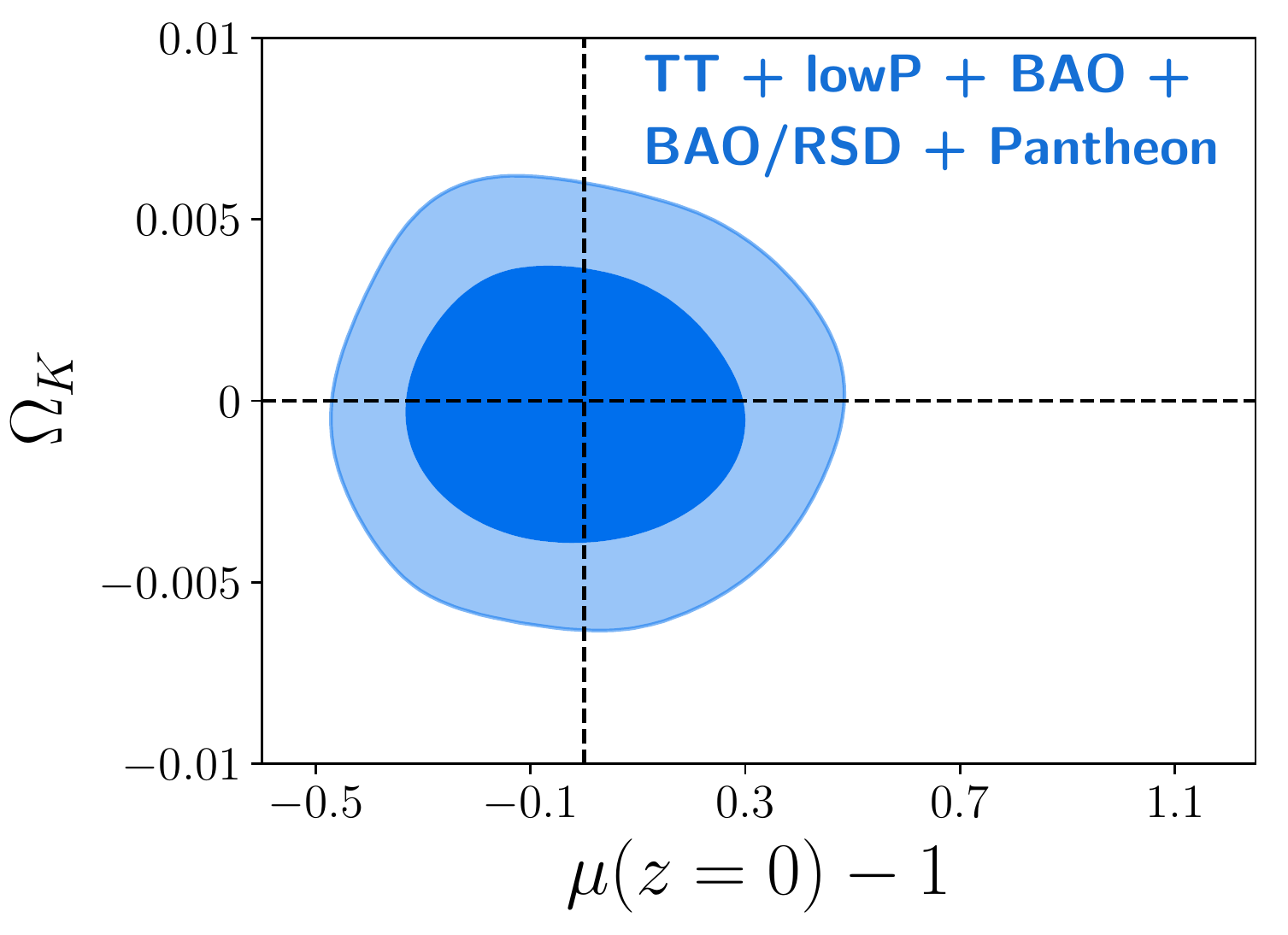}} 
  &  {\includegraphics[width = 5.8 cm]{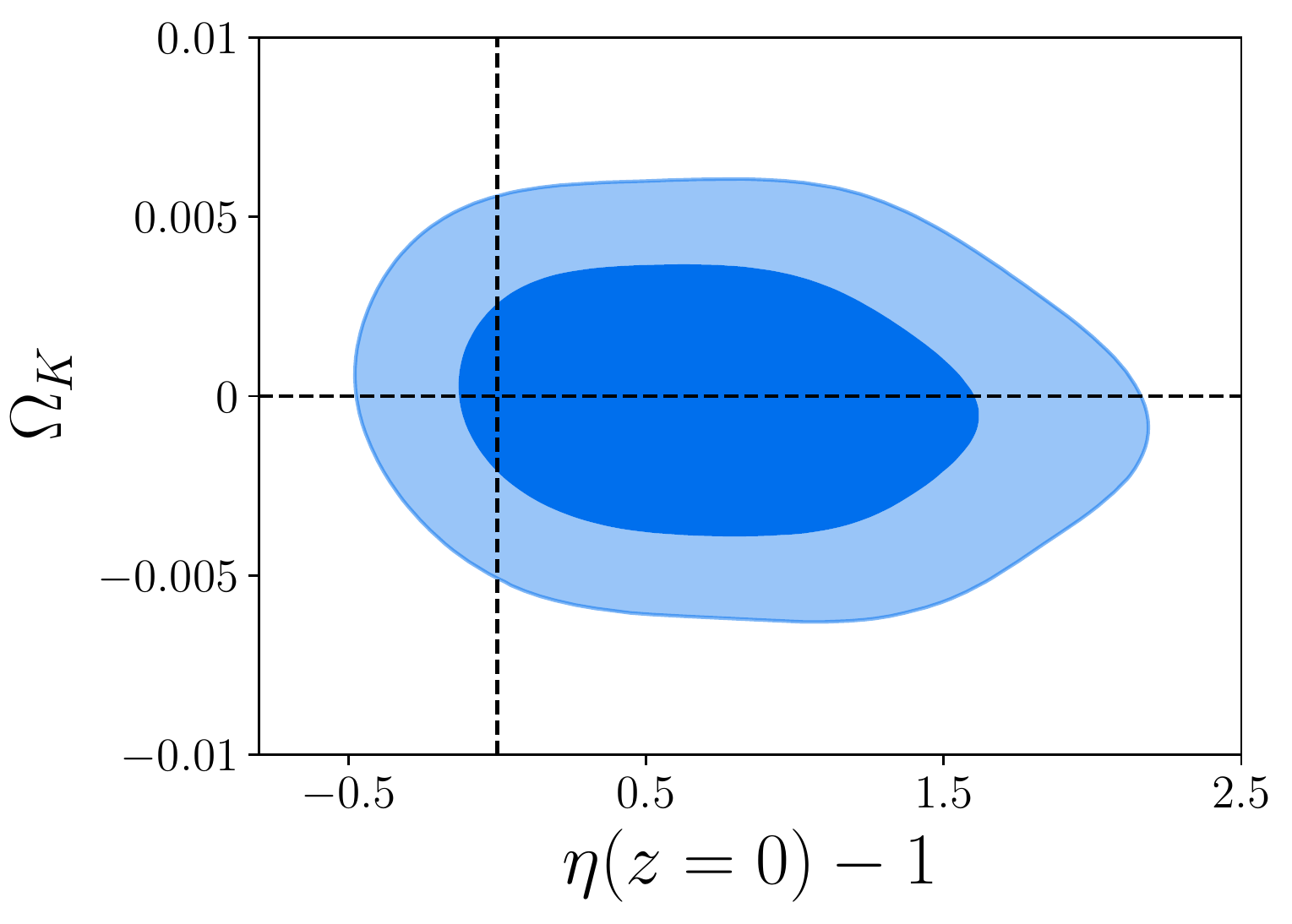}}
  &  {\includegraphics[width = 5.8 cm]{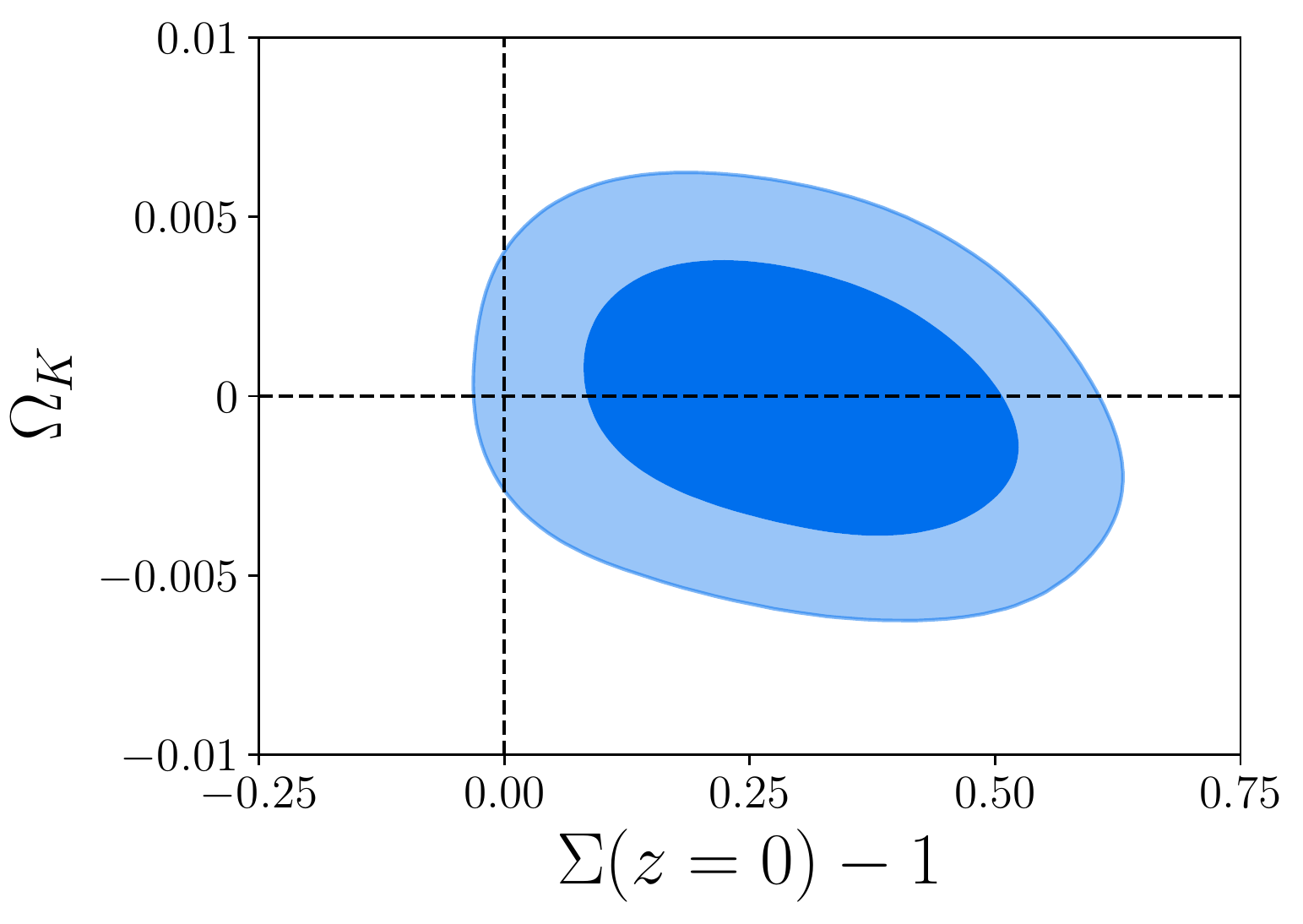}}
\end{tabular}
\end{center}

\caption{68\% and 95\% confidence contours for the parameters $\mu_0-1$, $\eta_0-1$ and $\Sigma_0-1$ with the addition of the extra feature of spatial curvature implemented in \texttt{ISiTGR}.}

\label{Plots_MGcurvature}

\end{figure}

When adding the scale dependence to MG parameters, we not only vary the parameters $E_{11}$ and $E_{22}$ for the time dependence, but also $c_1$, $c_2$ and $\lambda$ for the scale dependence. In Fig. \ref{Plot_scaledependence} we show the corresponding results. As we can see, for large scales the constraints become weaker because of the marginalization over the parameters $c_1$ and $c_2$.

Finally, we extend the results on this parametrization by adding spatial curvature to the $(\mu, \gamma)$ case. We show the corresponding constraints of the MG parameters in terms of $\Omega_k$ in Fig. \ref{Plots_MGcurvature}. Since we are not considering contributions from massive neutrinos in this case, we take into account the derive parameter $\Sigma-1$ at $z=0$. When the stress shear is negligible, the relationship $\Sigma = \frac{\mu}{2}(1+\eta)$ is satisfied.

%%%%%%%%%%%%%%%%%%%%%%%%%%%%%%%%%%%%%%%%%%%%%%%%%%%%%%%%%%%%%%
\subsubsection{Results for $(\mu,\Sigma)$ parametrization}

\begin{figure}[t!]
\begin{center}
 {\includegraphics[width=12cm]{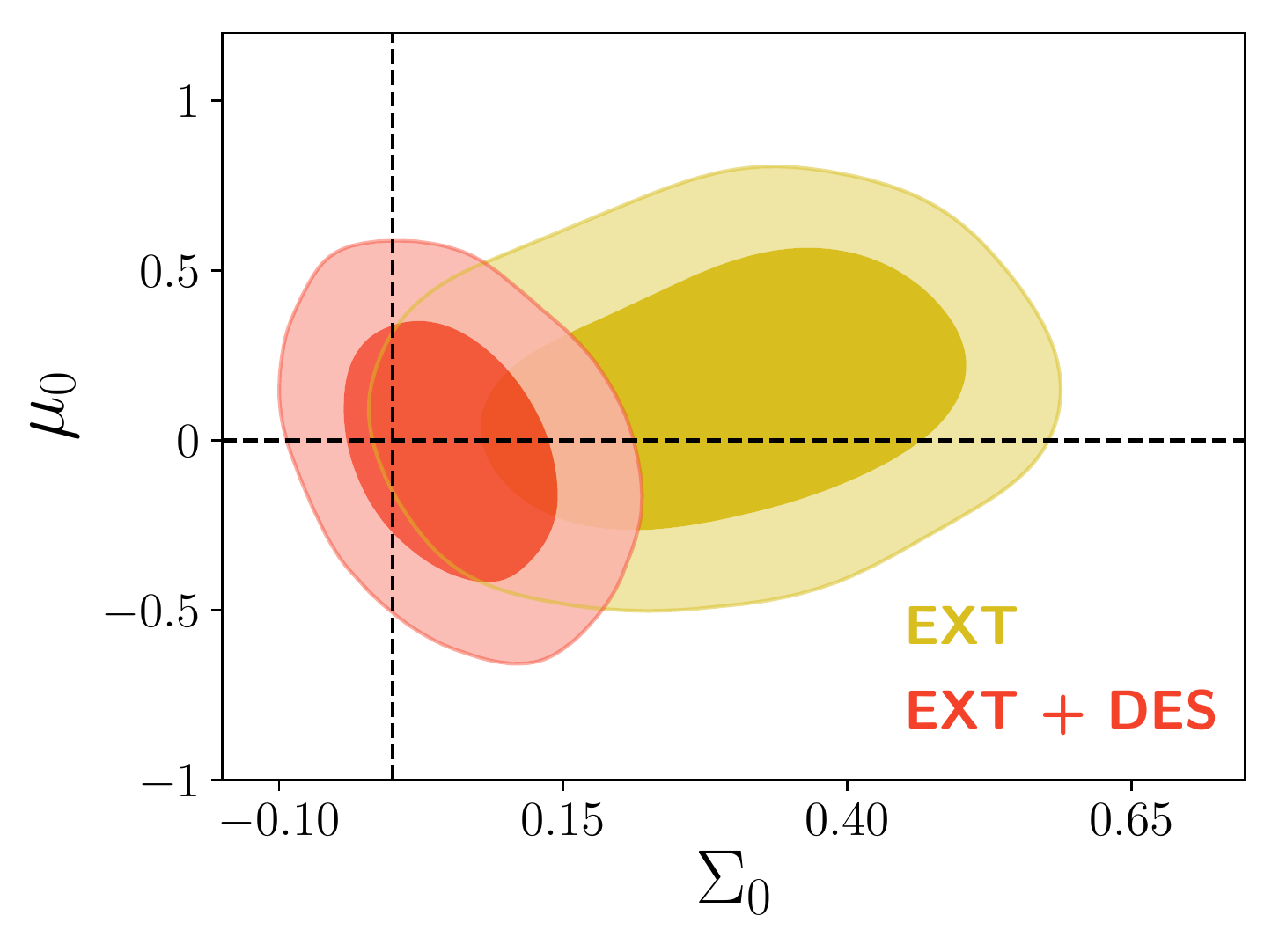}} 
\end{center}

\caption{68\% and 95\% confidence contour plots for $\mu_0$ and $\Sigma_0$ in the $(\mu,\Sigma)$ parametrization with DE-time evolution and no scale dependence. This is in agreement with results from DES 2018 \cite{DESMG2018} (Figure 3 there). 
Some small differences may be due to different priors, the likelihood/sampling methods or the process related to removing the non-linear data. We can see that adding the DES data to the EXT data set removes any tension with GR.}
\label{Plots_DES2018}
\end{figure}

\begin{table}[t!]
\begin{center}
 \begin{tabular} {p{2cm}  | >{\centering}m{4cm} >{\raggedleft\arraybackslash}m{2.5cm} }
\hline\hline \\ [-7pt]
 Parameter &  EXT  &  EXT + DES  \\[3pt]
\hline
 & &  \\[-8pt]
{$\mu_0          $} & $0.14\pm 0.27              $ & $0.03^{+0.30}_{-0.63}$ \\[5pt]

{$\Sigma_0       $} & $0.29^{+0.15}_{-0.13}      $ & $0.053^{+0.059}_{-0.066}   $ \\[5pt]
{$\Sigma m_\nu$} & $0.181^{+0.088}_{-0.12}    $ & $0.146^{+0.062}_{-0.11}    $ \\[5pt]

$\Omega_b h^2$ & $0.02244\pm 0.00021$ & $0.02242\pm 0.00021$\\[5pt]

$\Omega_c h^2$ & $0.1163^{+0.0018}_{-0.0016}$ & $0.1165^{+0.0015}_{-0.0013}$\\[5pt]

$100\theta$ & $1.04127\pm 0.00042$ & $1.04123\pm 0.00042$\\[5pt]

${\rm{ln}}(10^{10} A_s)$ & $3.041\pm 0.042$ & $3.082^{+0.033}_{-0.038}$\\[5pt]

$n_s$ & $0.9723^{+0.0050}_{-0.0057}$ & $0.9718\pm 0.0048$\\[5pt]

{$\tau$} & $0.058^{+0.021}_{-0.023}   $ &  $0.078^{+0.017}_{-0.020}   $ \\[5pt]
\hline\hline

\end{tabular}
\end{center}
\caption{Marginalized mean values of and 1-$\sigma$ errors on MG and cosmological parameters using DES+EXT data sets. The results are well within 1-$\sigma$ agreement with \cite{DESMG2018}.}

\label{TableResults2}

\end{table}

 We use here Eqs. (\ref{muEvolution}) and (\ref{SigmaEvolution}) to evolve $\mu$ and $\Sigma$, but set $\lambda=0$ to eliminate the scale dependence. Furthermore, we removed the non-linear data from DES measurements as suggested in \cite{Planck2015MG} and \cite{DESMG2018}, until we get $\Delta\chi^2 < 1$ where $\Delta\chi^2 \equiv (\boldsymbol{d}_{NL} - \boldsymbol{d}_{L})^T \boldsymbol{C}^{-1}(\boldsymbol{d}_{NL} - \boldsymbol{d}_{L})$ and $\boldsymbol{d}_{L}$ represents the vector non-linear data predictions and $\boldsymbol{d}_{L}$ represents the vector linear data predictions.

We let the neutrino mass vary freely and just for this case we consider the Planck likelihood file plik\_lite\_v18\_TT.CLIK instead of plik\_dx11dr2\_HM\_v18\_TT.clik. 
This makes it possible to compare some of our results to those of DES 2018 analysis of extended models \cite{DESMG2018}. 
We label the combination TT + lowP (2015) + CMBlens + BAO + BAO/RSD + Pantheon as EXT (for external data set). The priors on $\mu_0$ and $\Sigma_0$ were set as in \cite{DESMG2018}. In Fig. \ref{Plots_DES2018} and Table \ref{TableResults2}, we show our results from \texttt{ISiTGR} and find them in overall agreement with the DES 2018 analysis using \texttt{CosmoSIS} \cite{COSMOSIS} and \texttt{MGCAMB} \cite{MGCAMB2}.  Some small differences between the constraints may be due to other parameter priors (for example the neutrino mass), the likelihood/sampling method differences or the process related with removing the non-linear data.

\subsubsection{Results for different dark energy parametrizations}
\begin{figure}[t!]
\begin{center}
\begin{tabular}{ c c }
 {\includegraphics[width=8.5 cm]{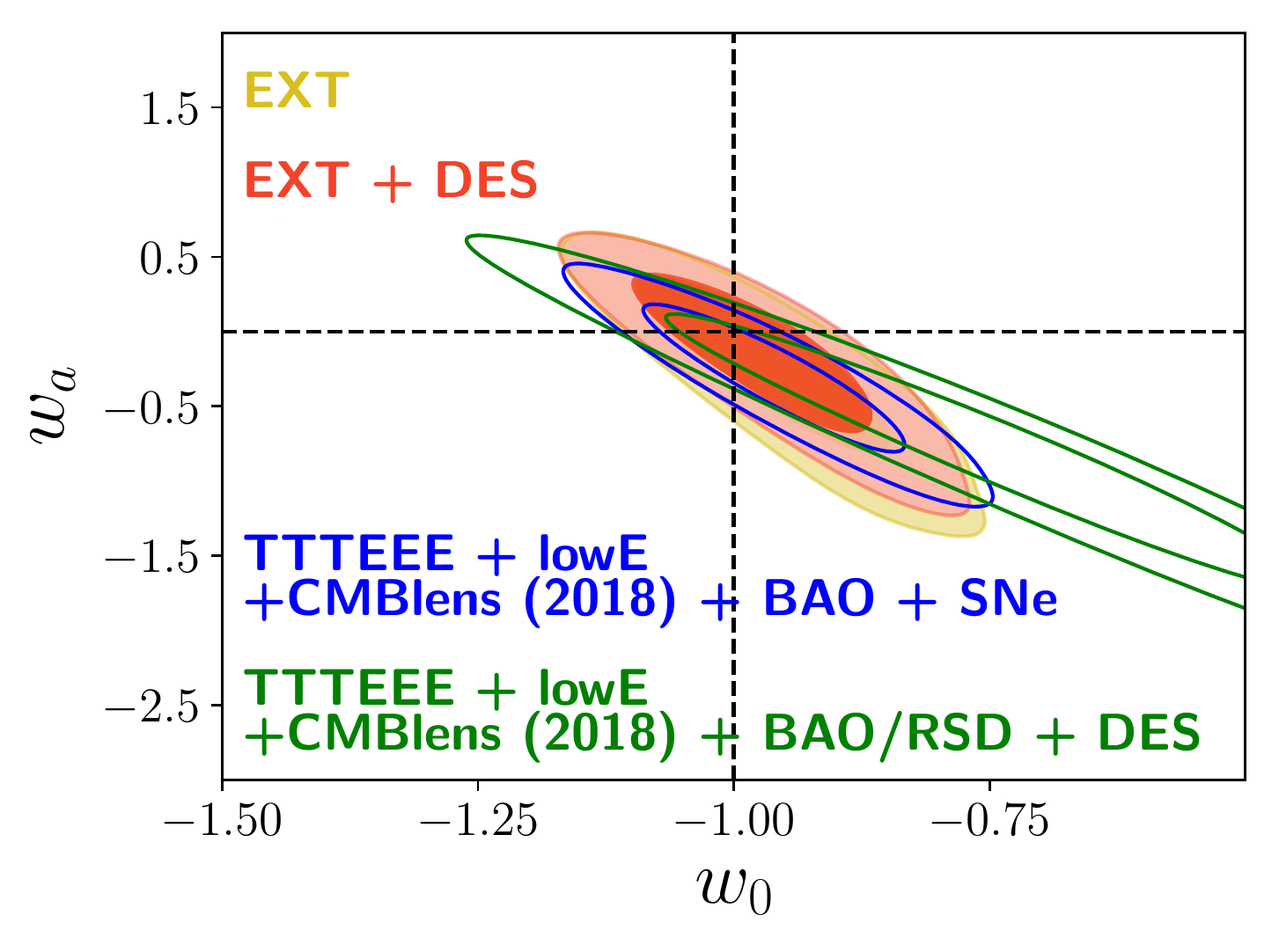}}
  &  {\includegraphics[width=8.5 cm]{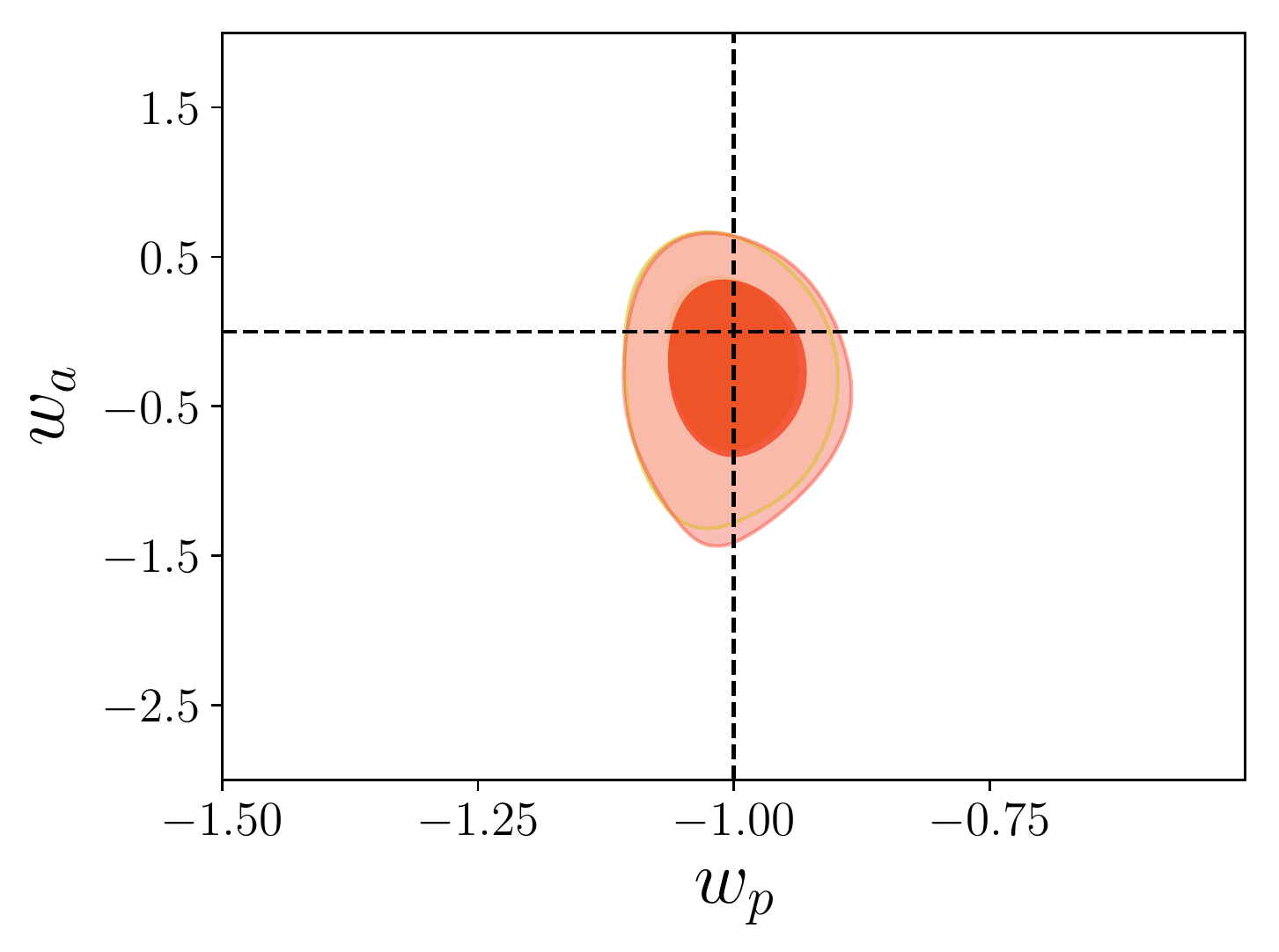}}
\end{tabular}
\end{center}
\caption{Left: 68\% and 95\% confidence contour plots for $w_0$ and $w_a$ for various data sets. We label the combination TT + lowP (2015) + CMBlens + BAO + BAO/RSD + Pantheon as EXT (for external data set) as done in DES 2018 so a direct comparison can be made. 
Right: Confidence contour plots for the pivot equation of state that uses $w_p$ and $w_a$. We can see that in the right-hand side plot, the parameters $w_p$ and $w_a$ are decorrelated. We used a pivot redshift of $z_p=0.21$ for EXT data, and the pivot redshift given in DES 2018, $z_p=0.20$, for EXT+DES data.}

\label{Plots_Pivot}

\end{figure}

\begin{table}[t!]
\begin{center}
\begin{tabular} { p{2cm}  | >{\centering}m{4cm}  >{\centering}m{4cm} >{\centering}m{4cm} >{\raggedleft\arraybackslash}m{3cm} } \hline\hline \\  [-6pt]
\scriptsize 
 Parameter &  EXT & EXT + DES & Planck-2018 + BAO +SNe & Planck-2018 + BAO/RSD + DES \\ \\[-8pt]
\hline
 & &  \\[-8pt]
$w_0$ & $-0.965^{+0.07}_{-0.08}$ & $-0.972^{+0.068}_{-0.081}$ & $-0.956\pm 0.082$ & $-0.74\pm 0.20 $ \\  [5pt]

$w_a$ & $-0.24^{+0.45}_{-0.27}$ & $-0.27^{+0.44}_{-0.29}$ &  $-0.32^{+0.34}_{-0.28}$ & $-0.83\pm 0.59$ \\ [5pt]

$w_p$ & $-1.005\pm 0.041$ & $-0.999\pm 0.043$ & - & - \\ [5pt]

$H_0$ & $67.84\pm 0.75$ & $67.82\pm 0.78$ &  $68.28\pm 0.83$ & $66.2^{+1.7}_{-1.9}$ \\ [5pt]

$\sigma_8$ & $0.812^{+0.016}_{-0.013}$ & $0.801^{+0.016}_{-0.014}$ &  $0.825\pm 0.011$ & $0.801^{+0.015}_{-0.017}$ \\ [5pt]

$S_8$ & $0.823\pm 0.014$ & $0.811\pm 0.014$ & $0.835\pm 0.011$ & $0.834\pm 0.013$ \\ [5pt]
\hline
\end{tabular}
\end{center}
\caption{68\% confidence limits for the DE parameters and cosmological parameters. The constraints are well within the 1-$\sigma$ level agreement with the constraints presented from \cite{DESMG2018,Planck2018}.}
\label{Table:DE-results}
\end{table}

\begin{figure}[t!]
\begin{center}
{\includegraphics[width=17cm]{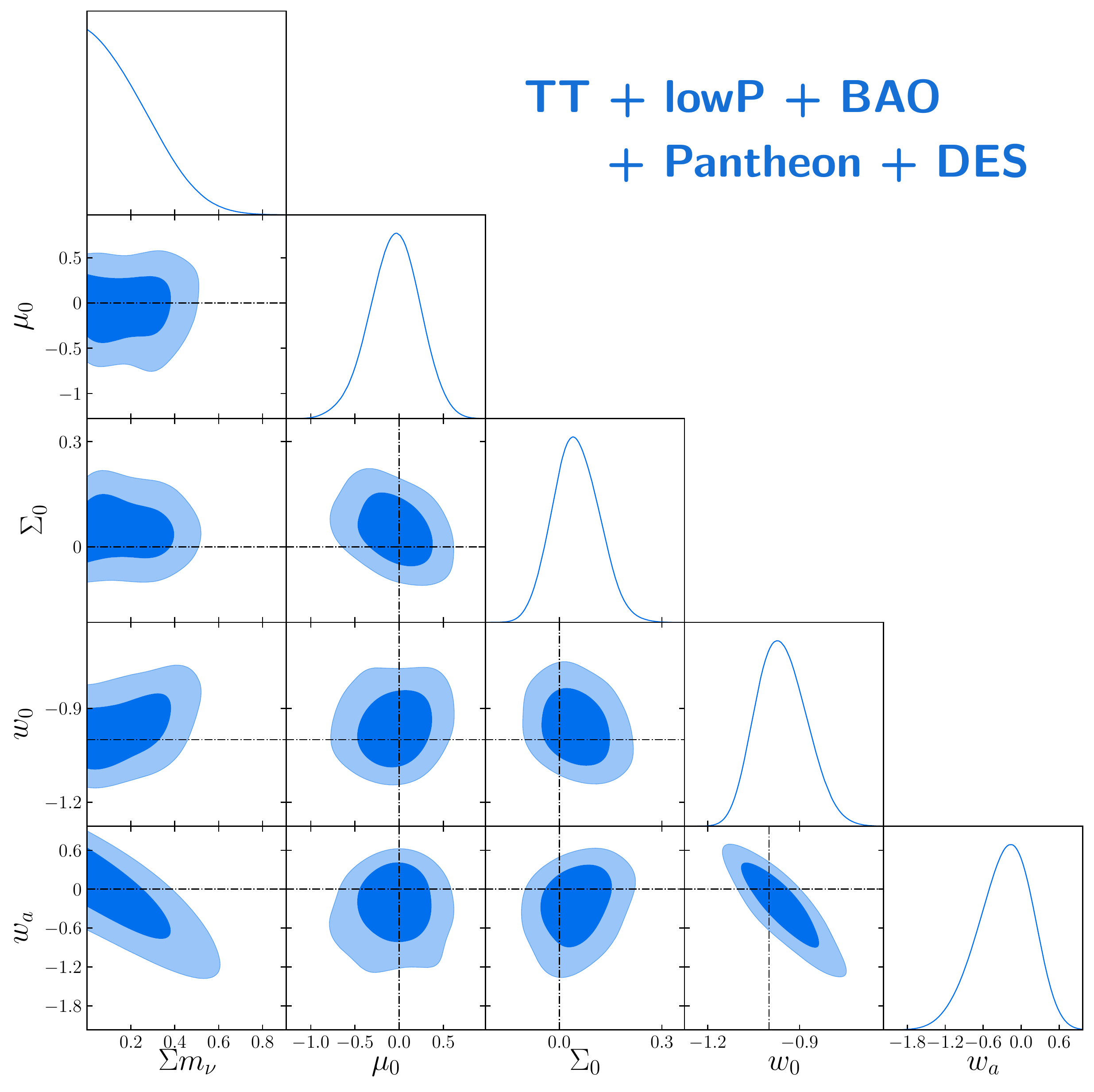}} 
\end{center}
\caption{68\% and 95\% confidence contours for the constraints obtained using the MG parameters $\mu_0$ and $\Sigma_0$ for the $(\mu,\Sigma)$ parametrization, combined with the dark energy equation of state $(w_0,w_a)$, and allowing the neutrino mass to vary. GR is found to be consistent with current data sets.}
\label{Plots_DEMGnu}
\end{figure}

In addition to the usual MG parameter approach, we also show our results for the new dark energy parametrizations implemented in \texttt{ISiTGR}. We present our results for dark energy parameter constraints in Table \ref{Table:DE-results}. The constraints shown for EXT and EXT + DES can be compared with the results obtained in \cite{DESMG2018}, while the constraints using Planck-2018 data should be compared with results in \cite{Planck2018}. The contraints are found within 1-$\sigma$ agreement  with these works. See Fig. \ref{Plots_Pivot} for the corresponding contour plots.

Furthermore, \texttt{ISiTGR} is able to combine these new dark energy parametrizations with MG parameters and massive neutrinos. We show in Fig. \ref{Plots_DEMGnu} the results obtained when these features are combine.

\subsubsection{Results for binning method for time and scale dependencies of MG parameters}
We derive results for the binning methods for the parametrizations $(\mu,\gamma)$ and $(\mu,\Sigma)$. Our results are shown in Figs. \ref{Plot_binning_mueta} and \ref{Plot_binning_muSigma}, as well as Table \ref{BinningResults}. Again, we find that binning methods complement very well functional methods. All results show that GR is consistent with current datasets. 
  
\begin{figure}[t!]
\begin{center}
\begin{tabular}{cccc}
   {\includegraphics[width=4.3cm]{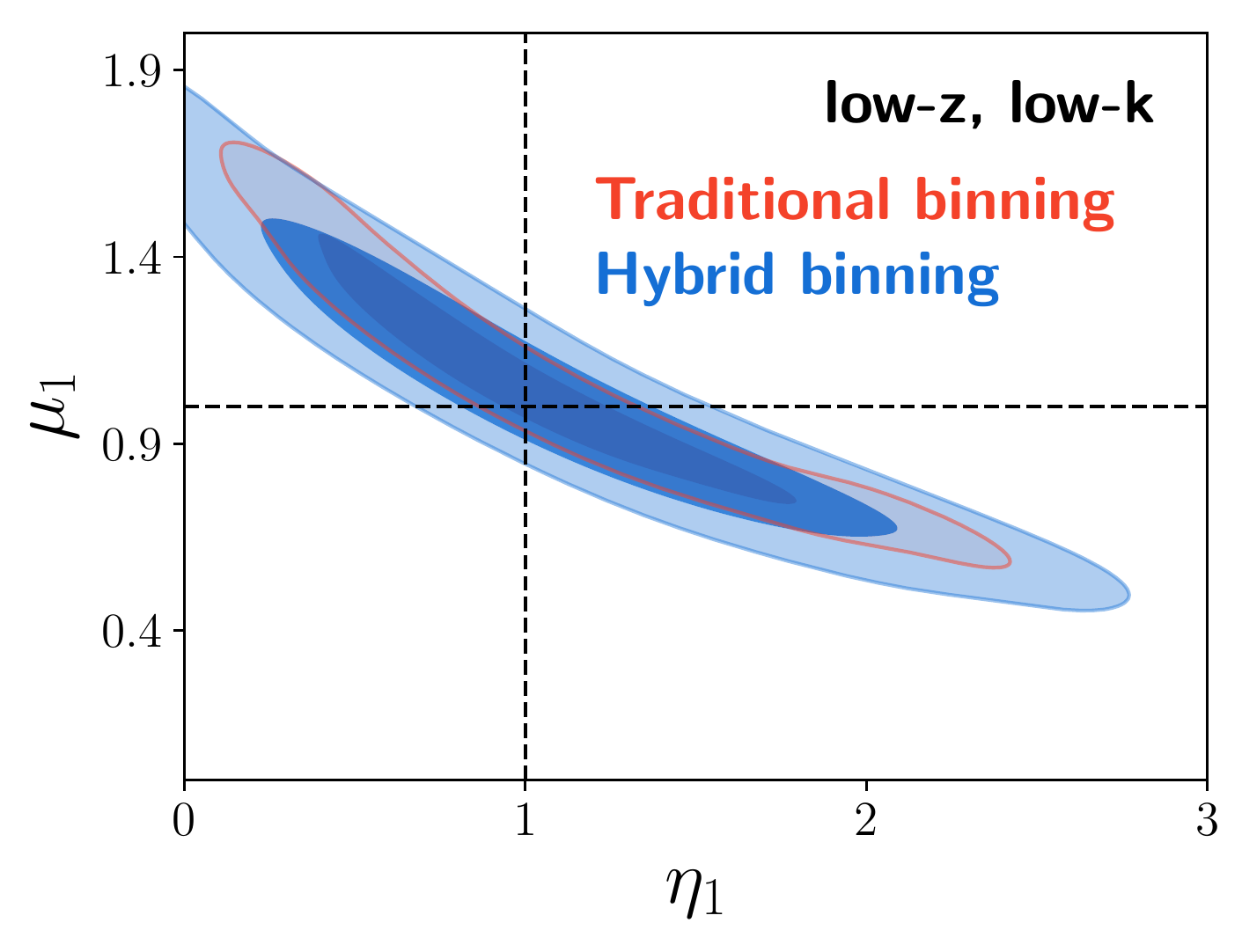}} 
  &  {\includegraphics[width=4.3cm]{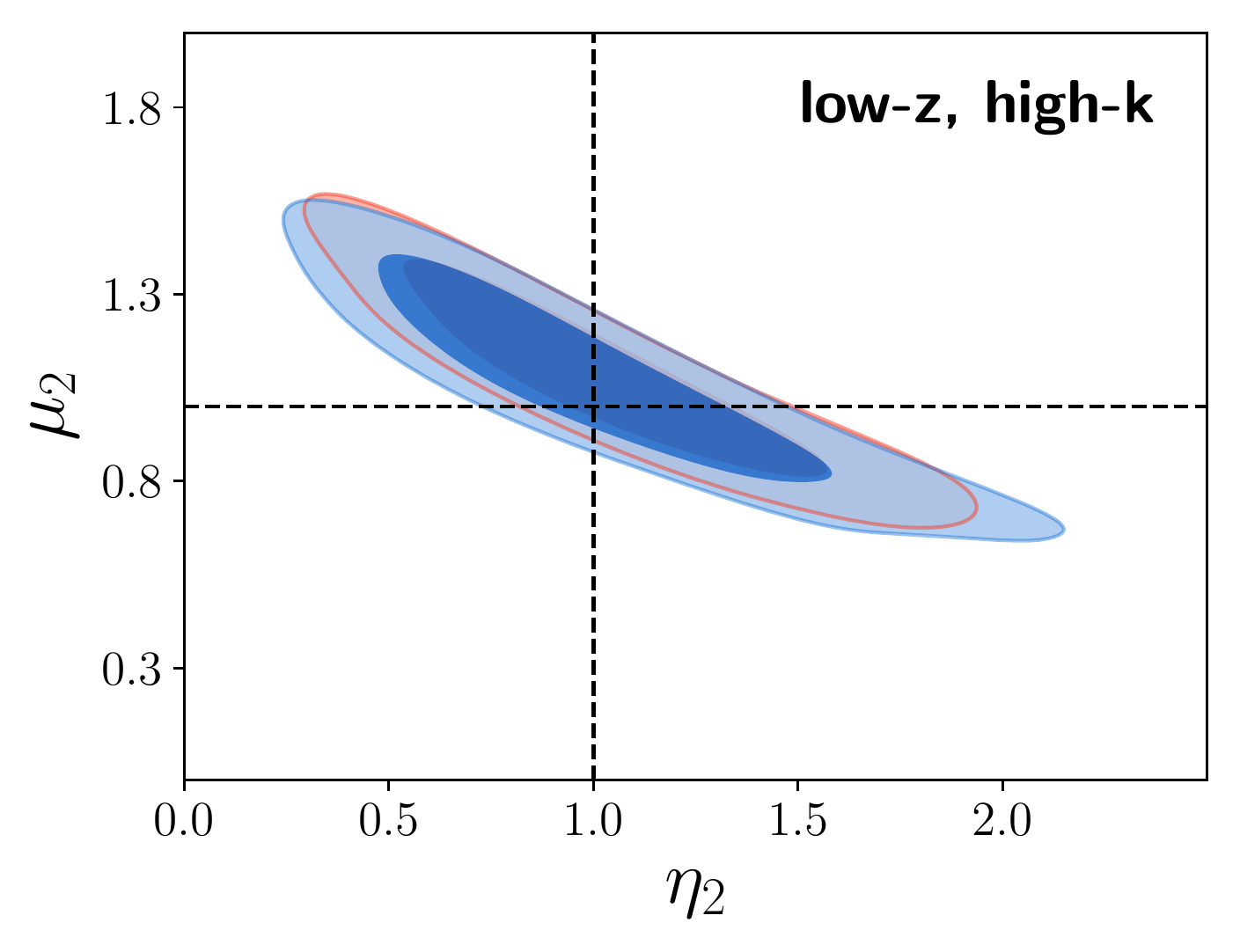}}  &
   {\includegraphics[width=4.3cm]{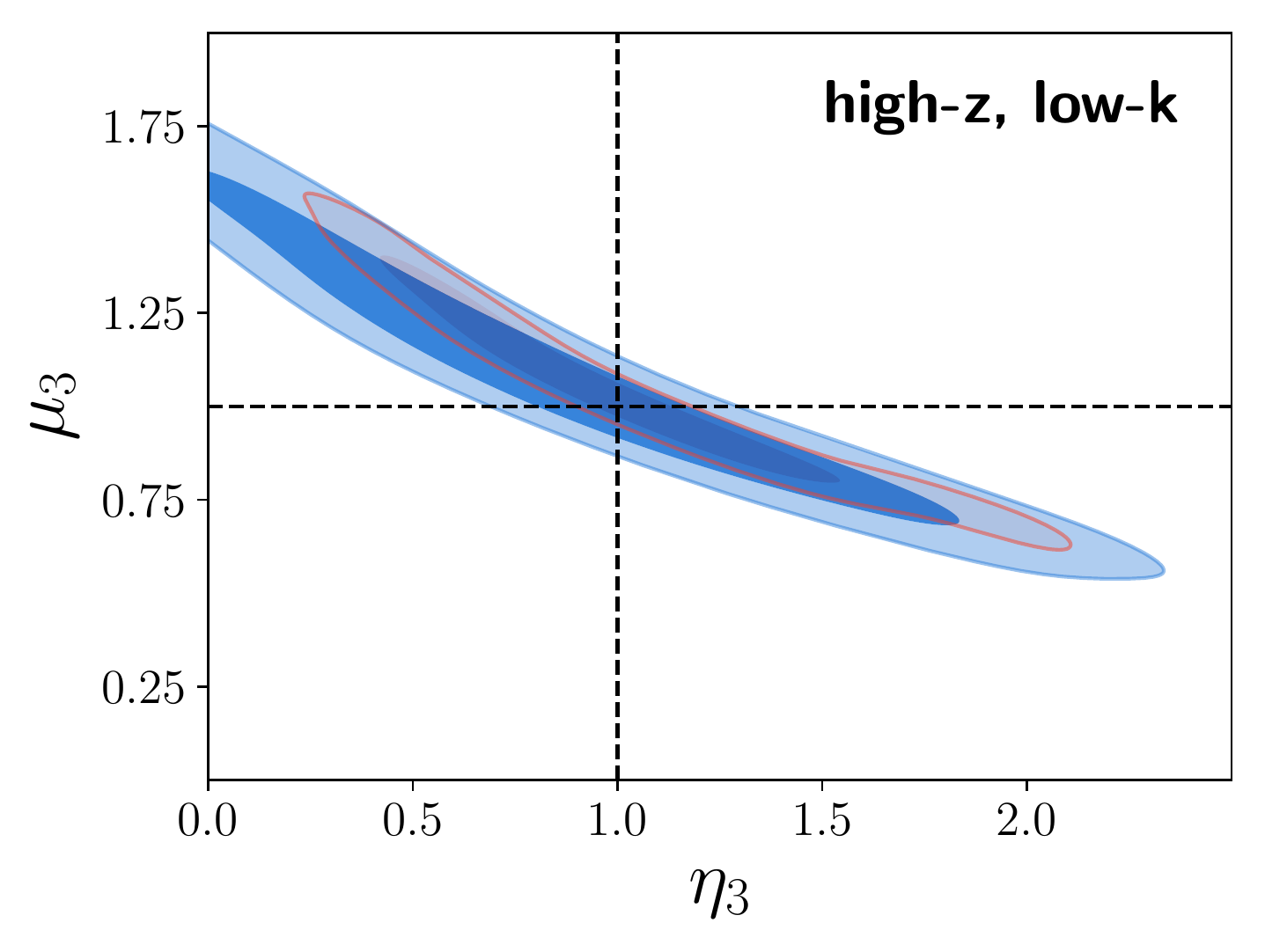}} &  {\includegraphics[width=4.3cm]{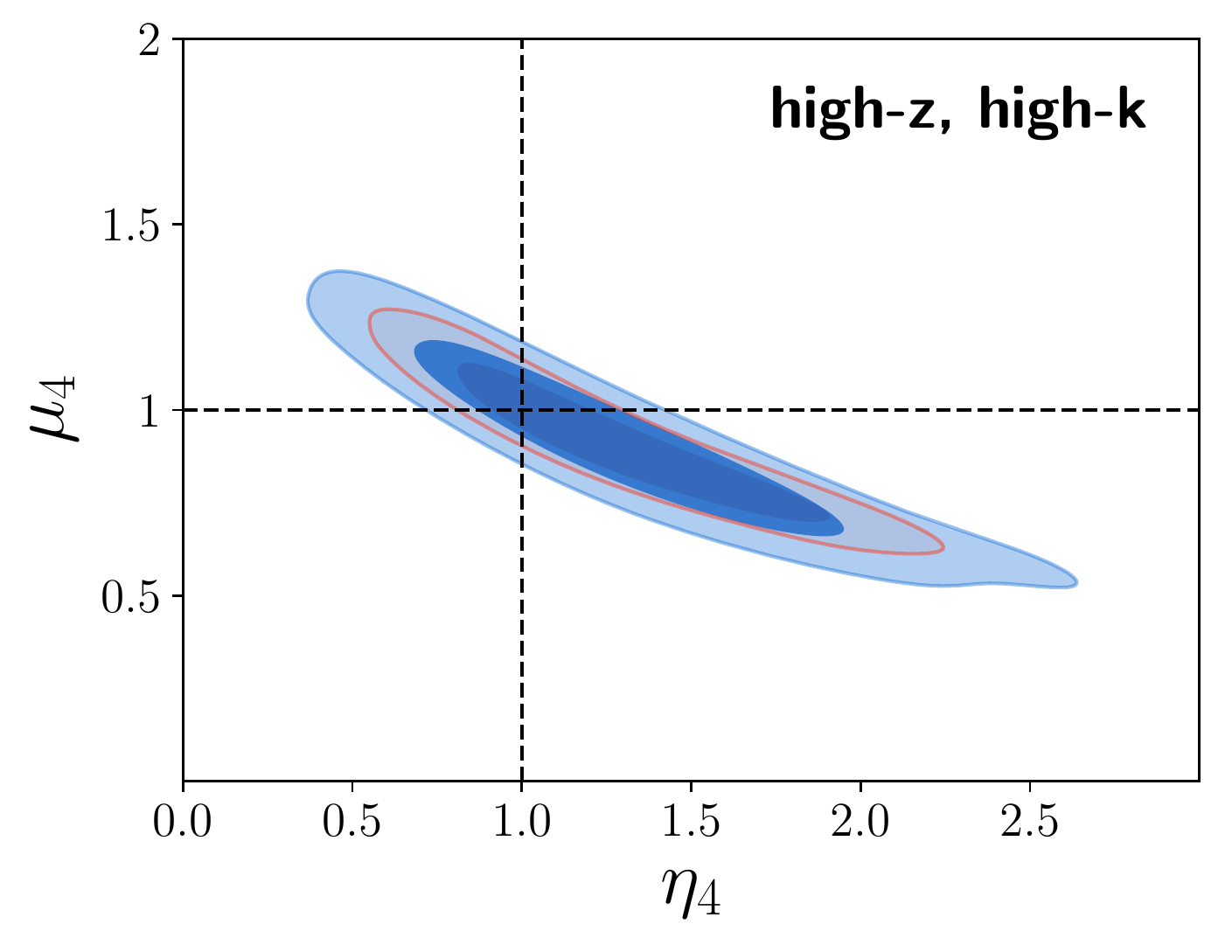}}
\end{tabular}
\end{center}
\caption{\label{Plot_binning_mueta} 68\% and 95\% confidence contours for both the traditional binning method and the hybrid binning method. To obtain the constraints we use the combination TT + lowP (2015) + Pantheon + BAO + BAO/RSD in a joint analysis. The z-bins are $0<z\leq 1$ and $1<z\leq 2$, with GR assumed for $z>2$. The k-bins are $k\leq 0.01$ and $k>0.01$. We plot the binning parameters for the $(\mu,\gamma)$ parametrization (in \texttt{ISiTGR}, $\gamma$ is referred to as $\eta$).}
\end{figure}

\begin{figure}[t!]
\begin{center}
\begin{tabular}{cccc}
 {\includegraphics[width=4.3cm]{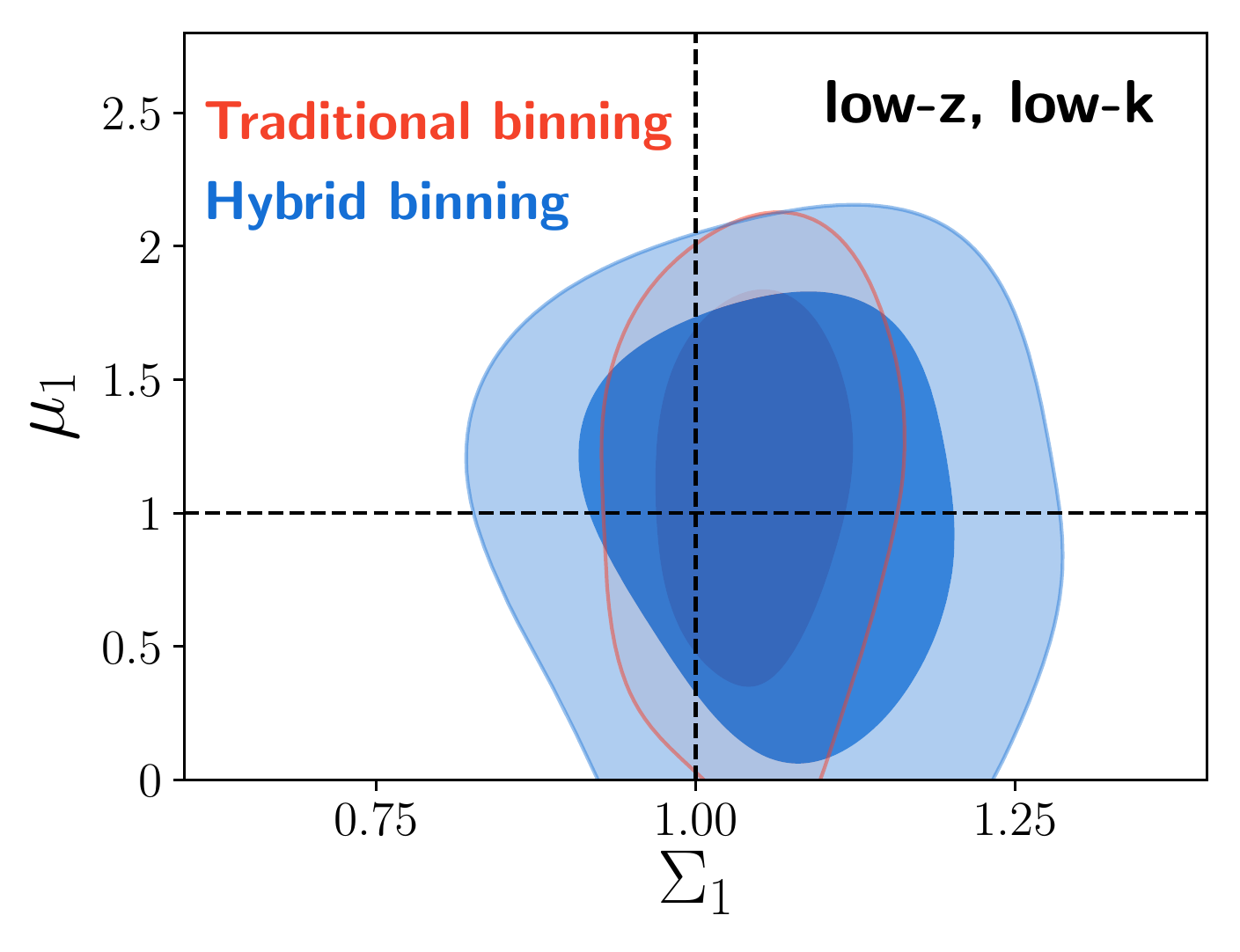}} 
  &  {\includegraphics[width=4.3cm]{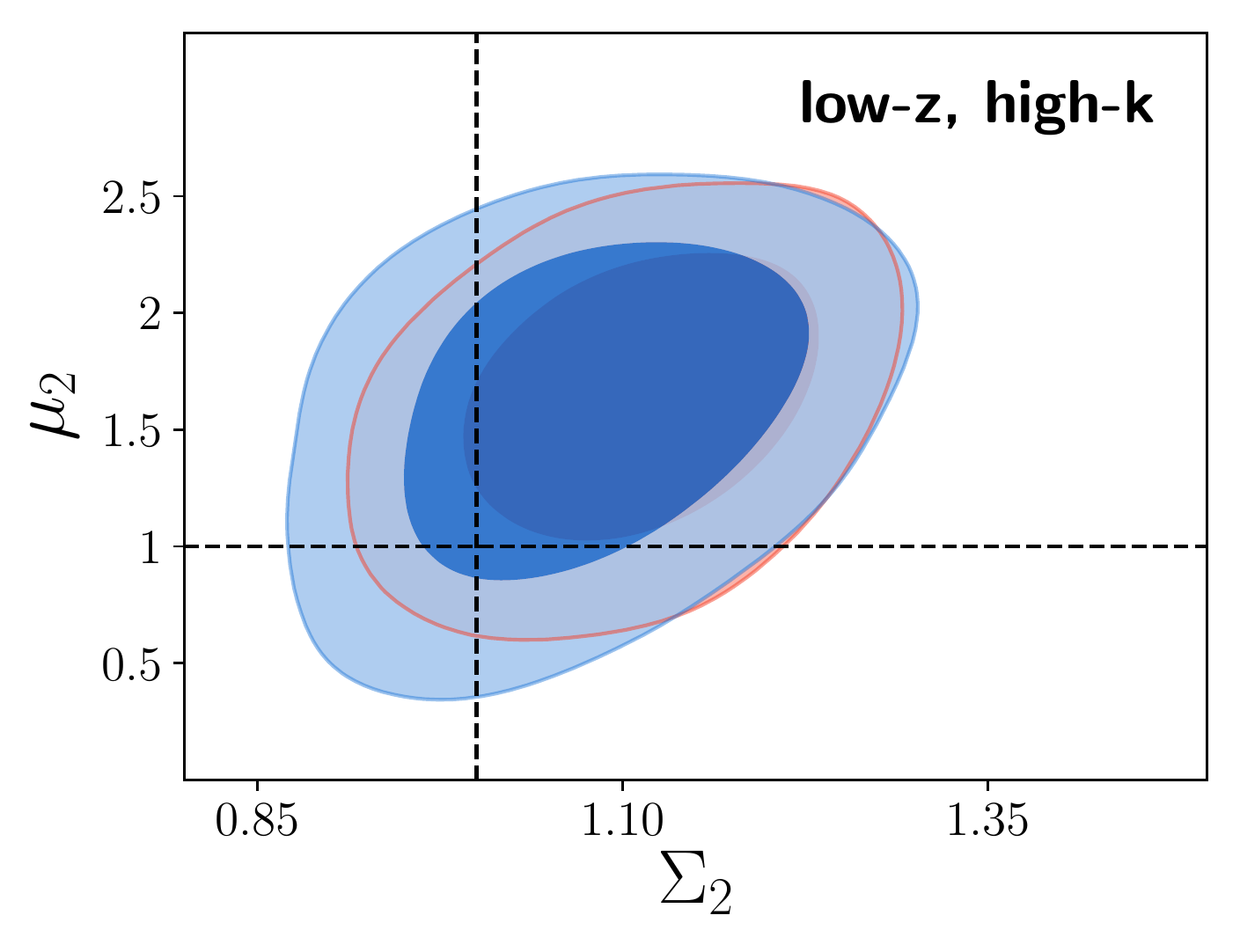}} &  {\includegraphics[width=4.3cm]{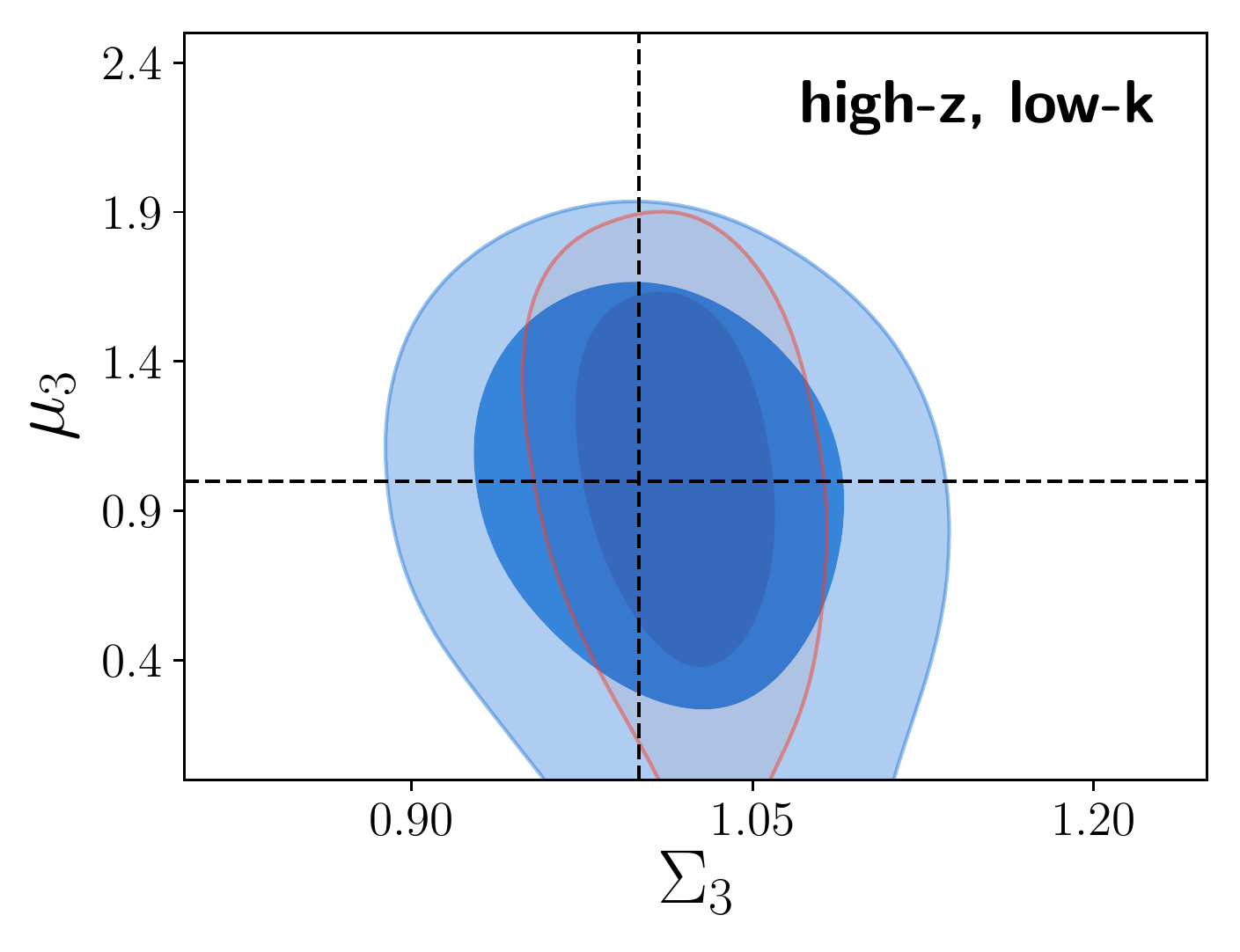}}  &  {\includegraphics[width=4.3cm]{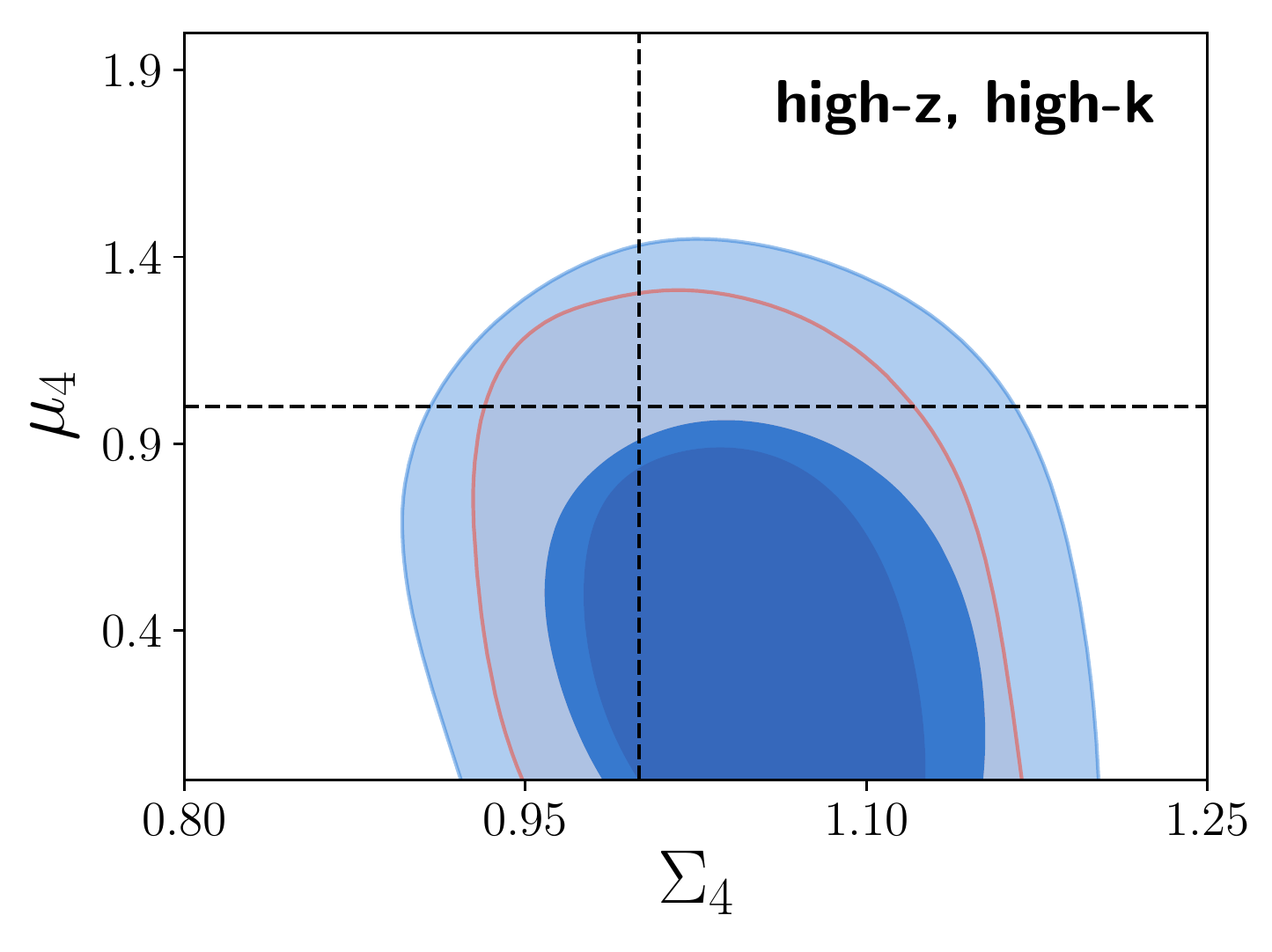}} \\
\end{tabular}
\end{center}
\caption{\label{Plot_binning_muSigma} 68\% and 95\% confidence contours for both the traditional binning method and the hybrid binning method, using the binning parameters for the $(\mu,\Sigma)$ parametrization. We use the combination TT + lowP (2015)+Pantheon+BAO+BAO/RSD in a joint analysis. The z-bins are $0<z\leq 1$ and $1<z\leq 2$ with GR assumed for $z>2$. Also, the k-bins are $k\leq 0.01$ and $k>0.01$.)}
\end{figure}

\begin{table}[t!]
\setlength{\tabcolsep}{2.5pt}
\begin{center}
\begin{tabular}{|c|c|c|c|c|c|c|c|c|}\hline\hline
\multicolumn{9}{|c|}{{ \bfseries Constraints for the MG parameters using the binning methods in the $(\mu,\gamma)$ parametrization }}\\ [3pt]\hline
\multicolumn{1}{|c|}{} & $\mu_1$ & $\mu_2$ & $\mu_3$ & $\mu_4$ & $\eta_1$ & $\eta_2$ & $\eta_3$ & $\eta_4$ \\ [3pt]
 Traditional binning & \, $1.06^{+0.52}_{-0.45}$ \,  & \,  $1.08^{+0.37}_{-0.35}$  \, &  \, $1.04^{+0.42}_{-0.38}$ \,  &  \, $0.91^{+0.28}_{-0.25}$ \,  & \,  $1.07^{+1.0}_{-0.90}$ \,  & \,  $1.03^{+0.72}_{-0.63}$  \, & \, $1.04^{+0.87}_{-0.73}$ & \, $1.30^{+0.69}_{-0.65}$ \\ [3pt]
 Hybrid binning & \, $1.02^{+0.55}_{-0.51}$ \,  & \,  $1.06^{+0.37}_{-0.37}$  \, &  \, $1.05^{+0.52}_{-0.47}$ \,  &  \, $0.92^{+0.36}_{-0.36}$ \,  & \, $1.2^{+1.1}_{-1.1}$\,  & \,  $1.03^{+0.81}_{-0.70}$  \, & \, $< 1.96$ & \, $1.30^{+0.92}_{-0.87}$ \\[3pt] \hline
\multicolumn{9}{|c|}{{\bfseries Constraints for the MG parameters using the binning methods in the $(\mu,\Sigma)$ parametrization }}\\ [3pt] \hline
\multicolumn{1}{|c|}{} & $\mu_1$ & $\mu_2$ & $\mu_3$ & $\mu_4$ & $\Sigma_1$ & $\Sigma_2$ & $\Sigma_3$ & $\Sigma_4$ \\ [3pt]
 Traditional binning & \, $1.11^{+0.84}_{-0.93}$ \,  & \, $1.63^{+0.77}_{-0.83}$  \, &  \, $1.01^{+0.75}_{-0.83}$ \,  &  \, $< 1.07$ \,  & \,  $1.043^{+0.096}_{-0.099}$ \,  & \,  $1.11^{+0.15}_{-0.16}$  \, & \, $1.017^{+0.056}_{-0.055}$ & \, $1.046^{+0.094}_{-0.10}$ \\ [3pt]
 Hybrid binning & \,  $1.05^{+0.82}_{-1.0}$ \,  & \, $1.58^{+0.85}_{-0.95}$ \, &  \, $0.98^{+0.76}_{-0.89}$ \,  &  \, $< 1.21$ \,  & \,  $1.07^{+0.19}_{-0.21}$ \,  & \,  $1.08^{+0.18}_{-0.18}$  \, & $1.01^{+0.11}_{-0.10}$ \, & \, $1.05^{+0.12}_{-0.13}$ \\ [3pt]\hline \hline
\end{tabular}
\end{center}
\caption{95\% confidence limits for the MG parameters in both the traditional binning method and the hybrid binning method. We present the results for the new parametrizations included in this new version of \texttt{ISiTGR}, the $(\mu,\gamma)$ and $(\mu,\Sigma)$ parametrizations. We found some small tension between the MG parameters and GR when looking at the 68\% confident limits. However, this disagreement disappears when considering the 95\% confidence limits.\label{BinningResults}}
\end{table}

\newpage
\newpage

{
%%%%%%%%%%%%%%%%%%%%%%%%%%%%%%%%%%%%%%%%%
\section{Summary and concluding remarks}
%%%%%%%%%%%%%%%%%%%%%%%%%%%%%%%%%%%%%%%%%
We have described in this paper the new version of the \texttt{ISiTGR} patch to test deviations from GR at cosmological scales using various data sets. This version now assembles the following capabilities:

\begin{enumerate}
    \item Dynamical dark energy parametrizations with a constant or time-dependent equation of state (also allowing the use of a pivot redshift to optimize constraints on its parameters).
    \item A consistent implementation through all formalism to account for anistotropic shear to model, for example, contributions from massive neutrinos.
    \item Spatially flat or curved backgrounds. 
    \item Multiple commonly used parametrizations of modified growth (MG) parameters to accommodate various types of data sets. 
    \item Functional and \emph{binned} time- and scale-dependencies of the MG parameters. 
\end{enumerate}  

As we are moving rapidly toward an era of precision cosmology with a plethora of incoming and future highly constraining data sets. It is important to have a software like \texttt{ISiTGR} that can constrain separately or  simultaneously various extensions to the standard cosmological model. 

Finally, We have also provided a number of results from applying \texttt{ISiTGR} to various available data sets and using various model extensions listed above. We find that GR is consistent with current data at cosmological scales. 

The code is made publicly available and ready to use at \url{https://github.com/mishakb/ISiTGR}.

\begin{acknowledgements}
We thank Shahab Joudaki, Weikang Lin and Eric Linder for useful comments on the manuscript. We thank Antony Lewis for clarifying some points about \texttt{CosmoMC} and Matteo Martinelli for clarifying some points about some of the MG results in \cite{Planck2015MG}. MI acknowledges that this material is based upon work supported in part by the U.S. Department of Energy, Office of Science, under Award Number DE-SC0019206 and the National Science Foundation under grant AST-1517768. CGQ gratefully acknowledges a PhD scholarship from the Mexican National Council for Science and Technology (CONACYT).

The authors acknowledge the Texas Advanced Computing Center (TACC) at The University of Texas at Austin for providing HPC resources that have contributed to the research results reported within this paper. URL: http://www.tacc.utexas.edu 
\end{acknowledgements}
}
\appendix{

\section{Spatially flat case}
\label{sec:flatcase}
For convenience and clarity we write explicit forms of various equations for the spatially flat case ($K=0$).

\subsection{Formalism and MG equations }

For $K = 0$ the various Poisson-like equations for the metric potentials, and their corresponding modifications with MG parameters, are

$\\(\mu,\gamma)$ parametrization:

\begin{equation}
k^2\Psi = -4\pi G a^2\mu(a,k) \sum_i\left[\rho_i\Delta_i+3\rho_i(1+w_i)\sigma_i\right]
\end{equation}
and
\begin{equation}
    k^2(\Psi-\gamma (a,k)\Phi) = -12 \pi G a^2\mu (a,k)\sum_i \rho_i(1+w_i)\sigma_i.
\end{equation}

$\\(\mu,\Sigma)$ parametrization:

\begin{equation}
 k^2\Psi = -4\pi G a^2\mu(a,k) \sum_i\left[\rho_i\Delta_i+3\rho_i(1+w_i)\sigma_i\right]
\end{equation}
and
\begin{equation}
    k^2(\Phi+\Psi)=-4\pi G a^2 \Sigma (a,k) \sum_i\left[2\rho_i\Delta_i + 3\rho_i(1+w_i)\sigma_i\right].
    \label{WeylFlat}
\end{equation}

$\\(Q,R)$ parametrization:

\begin{equation}
 k^2\Phi = -4\pi G a^2 Q(a,k) \sum_i\rho_i\Delta_i
\end{equation}
and
\begin{equation}
k^2(\Psi-R(a,k)\Phi) = -12\pi G a^2 Q(a,k) \sum_i\rho_i(1+w_i)\sigma_i.
\end{equation}

$\\(Q,D)$ parametrization:

\begin{equation}
 k^2\Phi = -4\pi G a^2 Q(a,k) \sum_i\rho_i\Delta_i
\end{equation}
and
\begin{equation}
k^2(\Phi+\Psi) = -8\pi G a^2 D(a,k) \sum_i \rho_i\Delta_i - 12\pi G a^2 Q(a,k)\sum_i\rho_i(1+w_i)\sigma_i.
\end{equation}
Note that $\Sigma$ is defined directly as a modification to the unmodified version of (\ref{WeylFlat}), while $D$ is defined in terms of $R$ and $Q$ as $D\equiv \frac{Q}{2}(R+1)$. $\Sigma$ can similarly be related to $\mu$ and $\gamma$, but the relation only takes the simple form $\Sigma=\frac{\mu}{2}(\gamma+1)$ in the limit of zero anisotropic shear $(\sigma_{i} = 0).$ It should also be noted that it is only in the case of zero shear that we have $\Sigma = D$.

\subsection{Implementation}
Following section \ref{sec:Implementation}, we can write equations for the synchronous gauge potentials for $K=0$ as
\begin{equation}
k^2(\eta-\mathcal{H}\alpha)  =-4\pi G a^2\sum_i \rho_i \Delta_i
\end{equation}
and
\begin{equation}
k^2(\dot{\alpha}+2\mathcal{H}\alpha-\eta) = -12 \pi G a^2\sum_i \rho_i(1+w_i)\sigma_i,
\end{equation}
where $\alpha = (\dot{h}+6\dot{\eta})/2k^2$, and again these potentials are related to the the Newtownian gauge potentials by
\begin{equation}
\Phi = \eta - \mathcal{H}\alpha
\end{equation}
and
\begin{equation}
\Psi = \dot{\alpha} + \mathcal{H}\alpha.
\end{equation} 
The expressions for $\dot{\eta}$ in the flat case become

$\\{(\mu,\gamma)}$
parametrization:
\begin{equation}
\begin{split}
\dot{\eta}_{(\mu,\gamma)} & = \frac{1}{2f_{\mu,\gamma}} \bigg\{ k\mu\gamma f_1 \sum_i q_i^{(S)}\hat{\rho_i} + \sum_i  \Delta_i\hat{\rho_i}[\mathcal{H}\mu(\gamma-1)-\dot{\mu}\gamma - \mu\dot{\gamma}] \\& 
 + 2\mu(1-\gamma)\sum_i\dot{\Pi_i}\hat{\rho_i} + k^2\alpha [-2(\mathcal{H}^2-\dot{\mathcal{H}})+\mu\gamma \sum_i\hat{\rho_i}(1+w_i)] \\&
 - 2[\mu\dot{\gamma}+\dot{\mu}(\gamma-1)]\sum_i\Pi_i\hat{\rho_i} + 2\mathcal{H}\mu\sum_i\Pi_i\hat{\rho_i}(\gamma - 1)(3w_i+2)
\bigg\},
\end{split}
\end{equation}
where
\begin{equation}
f_{\mu,\gamma} \equiv k^2+\frac{3}{2}  \mu\gamma\sum_i(1+w_i)\hat{\rho_i}. 
\end{equation}
$\\{(\mu,\Sigma)}$
parametrization:
\begin{equation}
\begin{split}
\dot{\eta}_{(\mu,\Sigma)} & = \frac{1}{2f_{\mu,\Sigma}} \bigg\{ k(2\Sigma-\mu) f_1 \sum_i q_i^{(S)}\hat{\rho_i} + [(\dot{\mu}-2\dot{\Sigma})+2\mathcal{H}(\Sigma-\mu)]\sum_i \Delta_i\hat{\rho_i} \\&
+ 2(\mu-\Sigma)\sum_i\dot{\Pi_i}\hat{\rho_i}  + 2[(\dot{\mu}-\dot{\Sigma})+\mathcal{H}(2\Sigma-\mu) -\mathcal{H}\mu]\sum_i\Pi_i\hat{\rho_i} \\&
+ 2\mathcal{H}(\Sigma-\mu)\sum_i\Pi_i(1+3w_i)\hat{\rho_i} + k^2\alpha \left[(2\Sigma-\mu)\sum_i(1+w_i)\hat{\rho_i}-2(\mathcal{H}^2-\dot{\mathcal{H}}) \right] 
\bigg\},
\end{split}
\end{equation}
where
\begin{equation}
f_{\mu,\Sigma}\equiv k^2+\frac{3}{2}(2\Sigma-\mu)\sum_i\hat{\rho_i}(1+w_i).
\end{equation}

$\\{(Q,D)}$
parametrization:

\begin{equation}
\begin{split}
\dot{\eta}_{(Q,D)} & = -\frac{1}{2f_Q} \bigg\{ 2k^2\alpha(\mathcal{H}^2-\dot{\mathcal{H}}) + [2 \mathcal{H}(D-Q)+\dot{Q}] \sum_i \Delta_i \hat{\rho}_i \\&
- k^2\alpha \sum_i  Q(1+w_i)\hat{\rho}_i -k Qf_1\sum_i q_i^{(S)}\hat{\rho}_i
\bigg\},
\end{split}
\end{equation}
where
\begin{equation}
    f_Q\equiv k^2+\frac{3}{2} Q\sum_i(1+w_i)\hat{\rho}_i.
\end{equation}
Note that $\dot{\eta}_{(Q,R)}$ can be obtained via the equation $R=\frac{2D}{Q}-1$.

Finally we need the derivative of the Weyl potential, $\dot{\Phi}+\dot{\Psi}$, to compute ISW effect contributions. During the derivation of the expressions for $\dot{\eta}$, $\dot{\Phi}$ was derived, so we just need expressions for $\dot{\Psi}$. Explicitly, in each case, these are given by

$\\(\mu,\gamma)$ or $(\mu,\Sigma)$ parametrizations:
\begin{equation}
\begin{split}
\dot{\Psi} & = -\frac{\dot{\mu}}{2k^2}\sum_i \left[\Delta_i\hat{\rho_i} + 2\Pi_i\hat{\rho_i}\right] + \frac{\mu}{2k^2}\sum_i \bigg\{ \mathcal{H}\Delta_i\hat{\rho_i} - 2\dot{\Pi_i}\hat{\rho_i} + k  f_1 q_i^{(S)} \hat{\rho_i} \\ &
+(1+w_i)\hat{\rho_i}[k^2\alpha f_1-3(\dot{\Phi}+\mathcal{H}\Psi)] + 2\mathcal{H}\Pi_i\hat{\rho_i}(1+3w_i) + 2\mathcal{H}\hat{\rho_i}\Pi_i \bigg\}.
\end{split}
\end{equation}
Note that this expression works for both of these parametrizations since it only involves $\mu$, however the explicit form of $\dot{\Phi}$ will depend on the particular parametrization used.

$\\(Q,D)$ or $(Q,R)$ parametrizations:
\begin{equation}
\begin{split}
\dot{\Psi}   = -\dot{\Phi}& - \frac{1}{k^2}\sum_i\bigg\{ \dot{D}\Delta_i\hat{\rho}_i+\dot{Q}\Pi_i\hat{\rho}_i - D\mathcal{H}\Delta_i\hat{\rho}_i - Q\mathcal{H}\Pi_i(3w_i+1)\hat{\rho_i} + Q\dot{\Pi}_i\hat{\rho}_i \\&
- 2 D\mathcal{H}\Pi_i\hat{\rho}_i- k Df_1 q_i^{(S)}\hat{\rho}_i + (1+w_i)\hat{\rho_i}[3D(\dot{\Phi}+\mathcal{H}\Psi) - k^2\alpha f_1 D]
\bigg\}.
\end{split}
\end{equation}
Note that this expression works for both of these parametrizations because the relation $R=\frac{2D}{Q}-1$ may be used.

\newpage
\section{Parametrization summary table}
\begin{table}[h!]
\begin{tabular}{|c|c|c|}
\hline\hline
set of MG parameters & MG first order perturbed equations & functional form of MG parameters \\ \hline
& & \\
& $(k^2-3K)\Psi = -4\pi G a^2 \mu(a,k)\left[\rho\Delta+3\frac{k^2-3K}{k^2}(\rho+p)\sigma\right]$ & $\mu=1+E_{11}\Omega_{DE}(a) \left[ \frac{1+c_1 \left( \lambda H/k \right)^2}{1+\left( \lambda H/k \right)^2} \right] $  \\
$(\mu,\gamma)$ &  &  \\
 & $k^2[\Phi-\gamma(a,k)\Psi]=12\pi G a^2\mu(a,k)(\rho+p)\sigma$ & $\gamma=1+E_{22}\Omega_{DE}(a) \left[ \frac{1+c_2 \left( \lambda H/k \right)^2}{1+\left( \lambda H/k \right)^2} \right] $  \\ 
 & & \\ \hline
 & & \\
 & $(k^2-3K)\Psi = -4\pi G a^2 \mu(a,k)\left[\rho\Delta+3\frac{k^2-3K}{k^2}(\rho+p)\sigma\right]$.  & $\mu=1+\mu_{0}\frac{\Omega_{DE}(a)}{\Omega_\Lambda} \left[ \frac{1+c_1 \left( \lambda H/k \right)^2}{1+\left( \lambda H/k \right)^2} \right] $ \\
$(\mu,\Sigma)$ & & \\
 & $k^2[\Phi+\Psi]=-4\pi G a^2\Sigma(a,k)\left[\frac{2}{1-3K/k^2}\rho\Delta+3(\rho+p)\sigma\right]$ & $\Sigma=1+\Sigma_{0}\frac{\Omega_{DE}(a)}{\Omega_\Lambda} \left[ \frac{1+c_2 \left( \lambda H/k \right)^2}{1+\left( \lambda H/k \right)^2} \right] $ \\ 
 & & \\ \hline
 & &  \\
 &  & $Q=1+Q_{0}\frac{\Omega_{DE}(a)}{\Omega_\Lambda} \left[ \frac{1+c_1 \left( \lambda H/k \right)^2}{1+\left( \lambda H/k \right)^2} \right] $ \\ 
 & &  \\
 & $(k^2-3K)\Phi = -4\pi G a^2 Q(a,k)\rho\Delta$ & $D =1+D_{0}\frac{\Omega_{DE}(a)}{\Omega_\Lambda} \left[ \frac{1+c_2 \left( \lambda H/k \right)^2}{1+\left( \lambda H/k \right)^2} \right] $ \\ 
$(Q,D)$ & & \\ \cline{3-1}
 & & \\
& $k^2[\Phi+\Psi] = -\frac{8\pi G a^2}{1-3K/k^2} D(a,k)\rho\Delta -12\pi G a^2 Q(a,k)(\rho+p)\sigma $  & $Q = [Q_0 e^{-k/k_c}+Q_{\infty} (1-e^{-k/k_c})-1]a^s+1$  \\ 
& & \\
& &  $D = [D_0 e^{-k/k_c}+D_{\infty} (1-e^{-k/k_c})-1]a^s+1$ \\
& & \\ \hline
 & &  \\
 &  & $Q=1+Q_{0}\frac{\Omega_{DE}(a)}{\Omega_\Lambda} \left[ \frac{1+c_1 \left( \lambda H/k \right)^2}{1+\left( \lambda H/k \right)^2} \right] $ \\ 
 & &  \\
 & $(k^2-3K)\Phi = -4\pi G a^2 Q(a,k)\rho\Delta$ & $R =1+R_{0}\frac{\Omega_{DE}(a)}{\Omega_\Lambda} \left[ \frac{1+c_2 \left( \lambda H/k \right)^2}{1+\left( \lambda H/k \right)^2} \right] $ \\ 
$(Q,R)$ & & \\ \cline{3-1}
 & & \\
& $k^2[\Psi-R(a,k)\Phi] = -12\pi G a^2 Q(a,k) (\rho+p)\sigma$ & $Q = [Q_0 e^{-k/k_c}+Q_{\infty} (1-e^{-k/k_c})-1]a^s+1$  \\ 
& & \\
& &  $R = [R_0 e^{-k/k_c}+R_{\infty} (1-e^{-k/k_c})-1]a^s+1$ \\
& & \\ 
\hline\hline
\end{tabular}
\caption{Summary table of the parametrizations used in ISiTGR.}
\end{table}
\newpage
\section{Flowcharts for the ISiTGR code}

\begin{figure}[h!]
\begin{center}
 {\includegraphics[width=16cm]{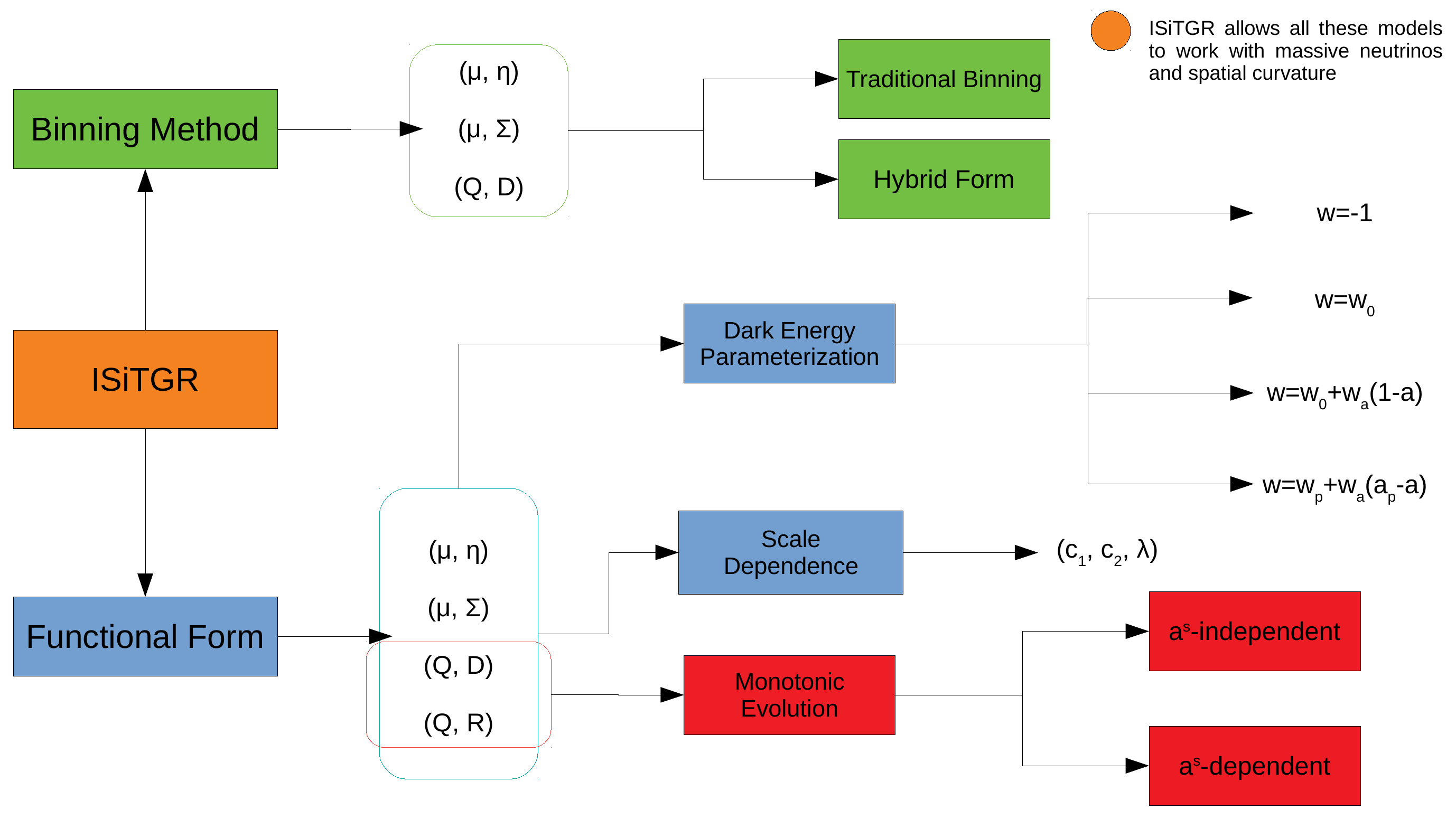}} 
\end{center}
\caption{Diagram that illustrates the various parametrizations of the \texttt{ISiTGR} code.}
\label{Plot_scaledependence}
\end{figure}

\begin{figure}[h!]
\begin{center}
 {\includegraphics[width=16cm]{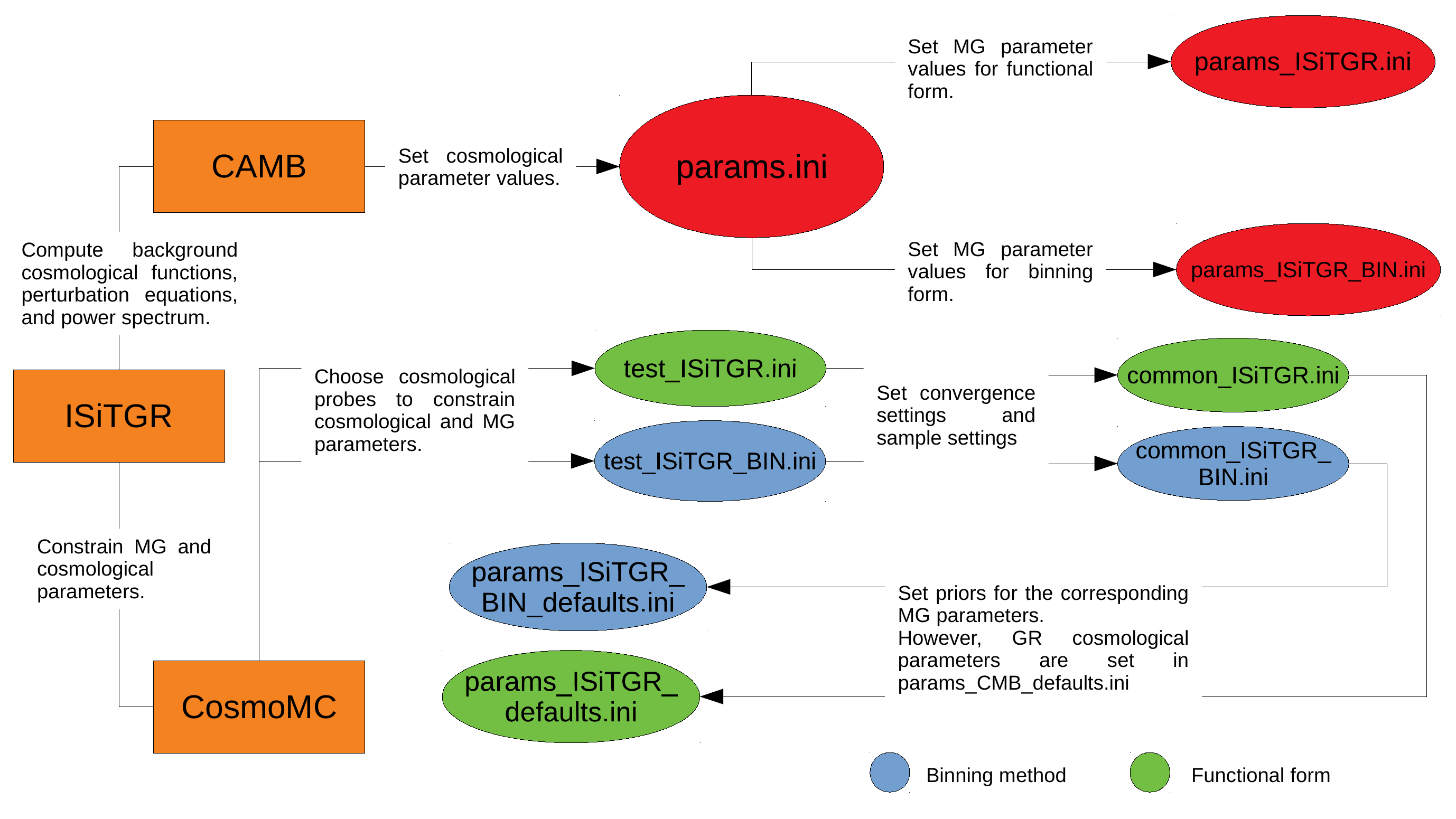}} 
\end{center}
\caption{Flowchart that shows the different files that need to be modified in order to use the \texttt{ISiTGR} patch. }
\label{Plot_scaledependence}
\end{figure}

}
\bibliography{ISiTGR_updated}

\end{document}